\documentclass[acmsmall]{acmart}

\usepackage{amsmath,amssymb,amsfonts}
\usepackage{algorithm} % for algorithm environment
\usepackage{algpseudocode} % for pseudocode
\usepackage{graphicx}
\usepackage{booktabs}
\usepackage{subcaption}

\usepackage{textcomp}
\usepackage{xcolor}

\usepackage{tikz}
\usetikzlibrary{shapes,arrows,positioning,fit,backgrounds}
\usepackage{tcolorbox}
\usepackage{todonotes}
\usepackage{enumitem}

\AtBeginDocument{%
  }

\acmJournal{TOSEM}
\acmVolume{}
\acmNumber{}
\acmArticle{}
\begin{document}

%\title{A Method to Estimate the Execution Times of Quantum Circuits on Quantum Computers}
\title{Understanding and Estimating the Execution Time of Quantum Circuits}

\author{Ning Ma}
\email{ning.ma@polymtl.ca}
\orcid{0009-0001-3148-7021}
\affiliation{%
  \institution{Polytechnique Montréal}
  \city{Montréal}
  \state{Québec}
  \country{Canada}
}

\author{Heng Li}
\email{heng.li@polymtl.ca}
\affiliation{%
  \institution{Polytechnique Montréal}
  \city{Montréal}
  \state{Québec}
  \country{Canada}
}
\authornote{Corresponding author}

\renewcommand{\shortauthors}{Ma and Li}

\begin{abstract}
Due to the scarcity of quantum computing resources, researchers and developers have very limited access to real quantum computers. Therefore, judicious planning and utilization of quantum computer runtime are essential to ensure smooth execution and completion of projects. Accurate estimation of a quantum circuit's execution time is thus necessary to prevent unexpectedly exceeding the anticipated runtime or the maximum capacity of the quantum computers; it also allows quantum computing platforms to make precisely informed provisioning and prioritization of quantum computing jobs.
In this paper, we first study the characteristics of quantum circuits' runtime on simulators and real quantum computers. Then, we introduce an innovative method that employs a graph transformer-based model, utilizing the graph information and global information of quantum circuits to estimate their execution time. We selected a benchmark dataset comprising over 1,510 quantum circuits, initially predicting their execution times on simulators, which yielded promising results with an R-squared value greater than 95\%. Subsequently, we applied active learning to select 340 circuit samples with a confidence level of 95\% to build and evaluate our approach for the estimation of circuit execution times on quantum computers, achieving an average R-squared value exceeding 90\%. 
Our approach can be integrated into quantum computing platforms to provide an accurate estimation of quantum execution time and be used as a reference for prioritizing quantum execution jobs. 
In addition, our findings provide insights for quantum program developers to optimize their circuits for reduced execution time. %, for example, by prioritizing one-qubit gates over two-qubit gates.
\end{abstract}




%% \received{20 February 2007}
%% \received[revised]{12 March 2009}
%% \received[accepted]{5 June 2009}

\maketitle
 
\section{Introduction}
\label{sec:introduction}
Over the last decade, there has been a significant transformation in quantum computing, evolving from experimental physics to cloud-accessible hardware (e.g., the IBM Quantum or Amazon Braket platforms), indicating a shift towards early-stage commercialization. This evolution has engaged a diverse audience, leading to research into practical quantum algorithms and gaining attention from academia, governments, and businesses. Notably, recent demonstrations of quantum utility~\cite{lundberg2017unified} suggest that quantum computers may eventually achieve comparable or better performance than classical counterparts for certain real-world tasks~\cite{zhong2020quantum}.

In quantum computing, the process of writing quantum programs\footnote{"Quantum programs" refers exclusively to circuit-level quantum programs in this paper.} and making a quantum circuit executable on a specific backend is akin to classical computing’s transformation of high-level code into low-level, machine-specific code. Designed initially without considering hardware constraints such as gate limitations or connectivity issues, a quantum circuit undergoes a series of compilation passes~\cite{Quetschlich2023mqtbench}. These passes progressively adapt the circuit to the target backend's restrictions. Given the complexity of these compilation problems, there is a variety of available approaches in various quantum SDKs and tools. These SDKs differ in the programming languages, gate sets, and quantum computer vendors they support~\cite{salm2021automating}. For example, vendor-agnostic SDKs such as pytket~\cite{sivarajah2020t} accommodate various languages and gate sets, whereas vendor-specific SDKs like IBM's Qiskit~\cite{javadi2024quantum} are restricted to their proprietary systems.

Executing a quantum circuit on a quantum computer involves exclusively using the quantum computer for a specific period, similar to the batch processing mode used in old mainframe systems. Users submit their designed and compiled quantum circuit to a pre-selected quantum computer, wait in a queue for processing, and finally the quantum computer executes the quantum circuit. The time taken to execute the quantum circuit is referred to as the "execution time". Given the scarcity and the substantial cost associated with quantum computing resources, the window available for using these state-of-the-art machines is exceptionally limited. Using IBM's Quantum Platform\footnote{https://quantum.ibm.com/} as an example, free users are limited to accessing two to three quantum computers per month, and usage is restricted to no more than 10 minutes. Any additional usage time is billed by the second and can be quite costly (\$1.6 USD per second at the time of writing). Exceeding time constraints
can lead to incomplete executions or cause significant costs, potentially delaying critical projects or incurring other severe consequences. Prudent planning and strategic utilization of quantum computer time are imperative to ensure the smooth execution and completion of projects. This underscores the necessity for an accurate estimation of quantum circuits' execution time. 
In addition, an accurate estimation of the execution time can help quantum computing platform providers (e.g., IBM Quantum Platform) with their provisioning and prioritization of quantum computing jobs.

Despite the crucial need for reliable execution time predictions for quantum circuits, current research in this area remains sparse. Based on our experience of using the IBM Quantum Platform, its provided execution time estimate is often inaccurate (more details are discussed in Section~\ref{sec:result}), failing to meet the demands of practitioners and researchers in the quantum computing field adequately. 
There are existing studies on estimating the execution time of classical programs~\cite{huang2010predicting, kwon2014mantis, meng2016nonlinear, bielecki2023estimation}. For instance, \citet{huang2010predicting} collected a set of program execution features of a classical program, such as \textit{loop counts} (the number of times a particular loop executed) and \textit{branch counts} (the number of times a particular conditional branch executed), and build polynomial regression models to predict its execution time. This method achieved an accuracy of over 93\% in predicting the execution time for the programs they tested. However, these approaches for classical programs do not apply to quantum programs, for three reasons: 1) quantum programs and classical programs are different in nature: classical programs describe the \textit{execution flows} of code statements, whereas quantum programs describe the \textit{structure} of a quantum circuit; 2) most approaches~\cite{huang2010predicting, kwon2014mantis, meng2016nonlinear} require tracing runtime information from the subject programs to collect features for model training or inference, whereas tracing quantum programs is not possible, at least for now, due to the famous no-cloning theorem~\cite{wootters1982single}; 3) these approaches build a model for each program~\cite{huang2010predicting, kwon2014mantis, bielecki2023estimation} or similar programs~\cite{meng2016nonlinear} as different programs have very different features (e.g., different loops or branches), but it is too expensive to collect sufficient training data for each quantum program or similar ones, as quantum computing resources are rather limited. 

To address this gap, this paper introduces an innovative methodology that harnesses graph transformer-based models to predict the execution times of quantum circuits, leveraging the topological graph data intrinsic to these circuits and their global characteristics. We have analyzed over 1,510 quantum circuits, ranging in size from 2 to 127 qubits, and executed them using different backends: two simulators and two real quantum computers. We organize our study by answering three research questions.

\begin{itemize}[leftmargin=*]
   \item \textbf{RQ1: How do the execution times of quantum circuits vary across different circuits and backends?}
\\ Understanding the execution times of quantum circuits is crucial for efficiently managing quantum computing resources. Intuitively, different quantum circuits with varying structures might be expected to have different execution times. However, execution times are influenced by multiple complex factors, including the number of qubits, gate counts, types of quantum gates, and the specific characteristics of quantum backends, etc. Due to these interacting factors, it might be possible that quantum circuits with distinct structural features could exhibit similar execution times. If the variations in execution times across different quantum circuits on different quantum backends are minimal, detailed execution-time estimation might lose its practical importance. Therefore, this research question aims to verify whether the differences in execution times among various quantum circuits are sufficiently substantial.
We observe that the execution times of different quantum circuits are noticeably different, whether on simulators or real quantum computers. In addition to these distributional differences, simple target-independent features (\emph{depth}, \emph{num\_qubits}, \emph{two-qubit gate count}, and \emph{num\_gates}) show nontrivial yet limited associations with execution time, and the associations differ across backends, indicating that single features are insufficient and motivating learned models that use backend-dependent features. Furthermore, the IBM Quantum Platform's estimation of execution times for quantum circuits on quantum computers is not sufficient, which leads us to the second and third research questions.

 \item \textbf{RQ2: How well can we estimate the execution times of quantum circuits on simulators, and what are the important predictors?}
\\ Quantum program developers typically use simulators to verify their programs before executing them on real quantum computers. Time-consuming simulation can cost significant computing resources (e.g., IBM Quantum's cloud-based simulators) and impact the productivity of quantum program development. Thus, understanding the simulation time of quantum circuits can help quantum developers better plan their simulation jobs (e.g., determining running a circuit for more or fewer shots) and help simulation service providers optimize the management of the simulation jobs.
In this RQ, we utilize various features, including features that capture the overall characteristics of a circuit (i.e., global features and features that capture the graph structure of a circuit (i.e., graph features), and construct machine learning models to estimate the execution time of quantum circuits on simulators. Our results demonstrate the precision of our automated estimation and highlight the influential features.

 \item \textbf{RQ3: How well can we estimate the execution times of quantum circuits on real quantum computers, and what are the important predictors?}
\\ Quantum computing resources are very limited currently and in the foreseeable future. Understanding the resource requirement (in terms of execution time) of quantum circuits before they execute is important for the optimal planning and allocation of quantum computing resources. Thus, this research question investigates the feasibility of creating a model that can accurately predict the execution times of quantum circuits on real quantum computers. Facing the limited quantum resources, we use active learning to collect the most representative training data. Our results show that we can achieve accurate predictions with a relatively small amount of data, substantially outperforming the estimation provided by the IBM Quantum Platform, and highlight the difference between the important predictors for the simulation time prediction and the prediction for real computer execution time.
\end{itemize}

Our work makes several important contributions. 

\begin{itemize}
\item We provided an in-depth understanding of the execution time of quantum circuits on both simulators and real quantum computers. 

\item We provided a novel and accurate approach for predicting the execution time of quantum circuits on both simulators and quantum computers. Our approach can help quantum program developers and resource providers better plan and allocate quantum computing resources.

\item We identified the influential features that explain the quantum circuit execution time, providing insights for quantum program developers to consider when developing efficient quantum circuits.

\end{itemize}

We share our replication package for future work to replicate or extend our work\footnote{https://github.com/mooselab/Quantum-Execution-Time-Prediction}.

\textbf{Paper organization.} 
Section~\ref{sec:related-work} discussed prior work related to our study. Section~\ref{sec:background} introduces the background knowledge of quantum computing, including quantum circuits, compilation, and execution. Section~\ref{sec:data-preparation} describes the data collection and preparation process. Section~\ref{sec:model} presents our model construction and the evaluation methods. Section~\ref{sec:result} presents and discusses our results for answering the research questions. Section~\ref{sec:discussion} gives a further discussion of our experiment. Section~\ref{sec:threats-to-validity} discusses the limitations of our study. Finally, Section~\ref{sec:conclusion} concludes our study and suggests directions for future research.

\section{Related Works} \label{sec:related-work}

As far as we know, there exist no prior work on understanding or estimating the execution time of quantum circuits. In this section, we discuss three categories of prior works that are closest to ours: quantum resource estimation, quantum circuit fidelity estimation, and classical software execution time estimation.

\subsection{Quantum resource estimation}
Quantum resource estimation offers a forward-looking solution for implementing key quantum algorithms by predicting necessary resources—such as qubits and runtime—based on future hardware projections, without running the actual programs. This approach allows consideration of quantum programs beyond current simulator and computer capabilities \cite{quetschlich2024utilizingresourceestimationdevelopment}.
QuRE~\cite{6657074} was among the pioneering efforts on quantum resource estimation. 
The authors developed a resource estimation methodology for diverse quantum technologies and algorithms, capable of automatically estimating metrics like qubit count, completion time, success probability, and gate counts for different physical gate types. Later on, another work has been undertaken to develop a framework for quantum resource estimation and evaluate the resource requirements for large-scale quantum applications, with a particular emphasis on the critical role of qubit parameters, aiming to expedite progress towards achieving practical quantum advantage~\cite{beverland2022assessing}. Another recent research provided resource estimates for quantum derivative pricing; it introduced a resource-efficient re-parameterization method, offered insights into resource demands for benchmark use cases, and outlined a roadmap for future quantum algorithm and hardware enhancements~\cite{Chakrabarti2021thresholdquantum}. 

Although tools such as QuRE have been developed to estimate quantum execution time, their applicable scenario is fundamentally different from our work. QuRE is designed as a theoretical resource estimation framework for fault-tolerant quantum computing. It estimates the total time to completion by modeling full-stack implementations of quantum algorithms, incorporating fault-tolerant error correction (e.g., Steane or surface codes), magic state distillation, tile-based layouts, and low-level physical gate scheduling. These estimates are based on hypothetical hardware models and assume the use of logical qubits constructed from many physical qubits, often requiring millions to billions of gates. In contrast, our work focuses on real-world near-term quantum computing, targeting NISQ devices. The execution time we estimate corresponds to the actual latency experienced by quantum circuits on realistic quantum backends, after compilation and transpilation. Our metric captures compilation-specific characteristics such as qubit mapping, SWAP insertion, and hardware-level gate duration. Thus, while both works use the term "execution time", the underlying definitions and contexts are not directly comparable.

Estimating resources often demands that end-users manually furnish extensive details about specific hardware characteristics, such as gate times and fidelities. Unlike previous research, our work does not require users to provide excessive hardware characteristic data or to have expert-level knowledge of quantum computing. Once a user writes a quantum program and selects a backend (either a simulator or a quantum computer) for execution, our methodology can automatically and accurately estimate the execution time of the quantum circuit on the chosen backend, making it highly accessible, efficient, and accurate. Another major difference is that our approach targets current backends (i.e., the simulators provided by Qiskit and quantum computers on the IBM Quantum Platform). This focus makes our model more practical and relevant for real-world applications.

\subsection{Quantum circuit fidelity estimation}
Fidelity is a crucial measure that describes the closeness of a quantum state or operation to a reference state or operation, quantifying the degree of similarity between quantum states or the precision in quantum information preservation~\cite{gilyen2022improved}. This measure is vital in practical quantum computing, where operations often face errors due to factors such as decoherence and imperfect gate implementations. High fidelity indicates minimal errors in quantum computations, crucial for reliable quantum computing. Recent strategies employ machine learning and deep learning to predict fidelity based on quantum circuit characteristics such as depth, and the type and number of quantum gates~\cite{vadali2024quantum}. Additionally, machine learning classification methods are used to select the most suitable backend for compiling a specific quantum circuit to maximize fidelity~\cite{quetschlich2023compiler, quetschlich2023predicting}.

Despite the existence of these studies on estimating the fidelity of quantum circuits on backends such as simulators and quantum computers, there is a notable gap in research utilizing the characteristics of quantum circuits to predict execution times on quantum computers. To address this deficiency, this paper aims to leverage both the characteristics of quantum circuits and their graph information to estimate the execution times, an equally important and meaningful area of research.

\subsection{Classical software execution time estimation}
Current research has created methodologies that link the execution times of computer programs to their code features, enabling accurate predictions of execution times for conventional programs. ~\citet{huang2010predicting} presents the SPORE (Sparse Polynomial Regression) methodology for predicting program execution time using a sparse, non-linear model based on features extracted from program execution data.~\citet{kwon2014mantis} developed Mantis, a framework designed to predict the execution time of Android applications. By leveraging program analysis and machine learning, Mantis identifies key program features strongly correlated with computational resource consumption (CRC) and automatically generates executable code snippets to evaluate them.~\citet{meng2016nonlinear} delves into an innovative nonlinear approach to estimate the worst-case execution time (WCET) during the programming phase. This approach uses historical WCET data from programs with similar object codes, including variations of those programs, to make predictions. However, practical challenges such as restricted access to source code due to licensing restrictions often emerge. \citet{bielecki2023estimation} develop models that use program metadata and runtime environment parameters as explanatory variables, yet treat the program code itself as a black box. These models primarily aim to estimate the average-case execution time (ACET).

Inspired by these studies, in this paper, we also utilize characteristics of quantum circuits (such as the number of quantum bits and the depth of the circuits) as well as the graph structures of quantum circuits along with the features of the corresponding backends (simulators and quantum computers) to predict the execution time of a quantum circuit on a specific backend.
\section{Background}
\label{sec:background}

\subsection{Quantum circuits}
\begin{figure} [h]
    \centering
    \includegraphics[width=0.40\linewidth]{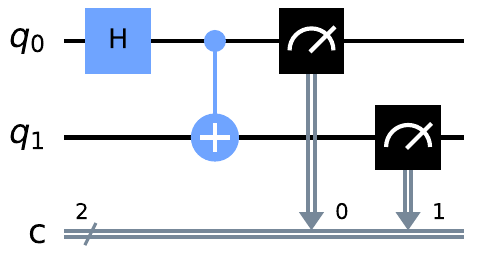}
    \caption{An example of a Quantum Circuit}
    \label{fig:An example of a Quantum Circuit}
\end{figure}

Fig.~\ref{fig:An example of a Quantum Circuit} shows a simple quantum circuit that we developed using IBM's open-source software development toolkit, Qiskit. 
A quantum circuit typically consists of a combination of quantum bits (qubits), classical bits, quantum gates, and measurement gates.
In the example shown in Fig.~\ref{fig:An example of a Quantum Circuit}, \( q_0 \) and \( q_1 \) represent two qubits, whereas \( c \) represent two classical bits. Unlike classical bits, each qubit can exist simultaneously in both $|0\rangle$ and $|1\rangle$ states.

The circuit includes two quantum gates which manipulate the states of the qubits, represented as follows:

\begin{itemize}
    \item The blue box with "H" symbol: Hadamard gate, applied to \( q_0 \).
    \item The blue box with "+" symbol: CNOT gate, with \( q_0 \) as the control qubit and \( q_1 \) as the target qubit.
\end{itemize}

At the end of the circuit, there are two black boxes indicating the measurements of the two qubits \( q_0 \)  and \( q_1 \) respectively. The measurements collapse the quantum states of the qubits into classical states and store the results in the two classical bits. 

\subsection{Quantum circuit compilation}\label{sec:background:compilation}

Like the classical domain, implementing a quantum algorithm on a real backend involves transforming it into a format that meets the specific hardware constraints~\cite{quetschlich2023mqt}. This transformation encompasses four stages: the algorithmic level, target-independent level, target-dependent native gates level, and target-dependent mapped level.

Quantum algorithms are typically designed and trialed at a hardware-neutral stage, known as the algorithmic level~\cite{quetschlich2023mqt}. Here, they are expressed as quantum circuits made up of high-level elements, free of restrictions related to specific gate-sets or hardware designs.

The initial step in applying a conceptual quantum algorithm to a specific problem is to create and refine these high-level components, disregarding the intended hardware. This process is akin to the function of a classical compiler and includes optimizing and simplifying the representation, now termed the target-independent level~\cite{quetschlich2023mqt}.

Modern quantum backends have specific limitations, necessitating a target-dependent compilation phase. Backends offer a limited range of native gates, usually an entangling gate and a set of single-qubit gates. Circuits must therefore be adapted to this native gate-set and optimized to minimize additional complexity. This adaptation is known as the target-dependent native gates level~\cite{quetschlich2023mqt}.

Additionally, current backends, particularly those using superconducting qubits, have restricted qubit connectivity. This requires mapping the circuit's logical qubits to the backend's physical qubits, ensuring that multi-qubit gates apply only to directly connected qubits. This dynamic mapping, followed by optimization to reduce compilation overhead, is defined as the target-dependent mapped level~\cite{quetschlich2023mqt}.

Finally, the circuit is prepared for execution on the quantum computer. It involves scheduling the gates, calibrating, and converting the circuit into machine code~\cite{quetschlich2023mqt}. The final binary code is then executed on the backend to perform the quantum computation.

\subsection{Quantum execution and execution time}

\begin{figure}[htbp]
    \centering
    \begin{subfigure}[htbp]{1\textwidth}
        \includegraphics[width=\textwidth]{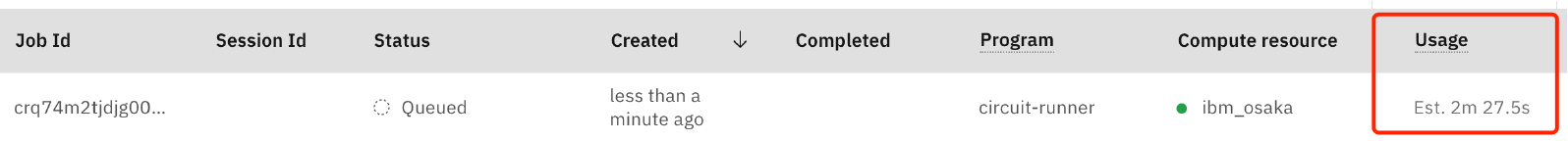}
        \caption{Estimated Execution Time}
        \label{fig:estimated}
    \end{subfigure}
    \hfill
    \begin{subfigure}[htbp]{1\textwidth}
        \includegraphics[width=\textwidth]{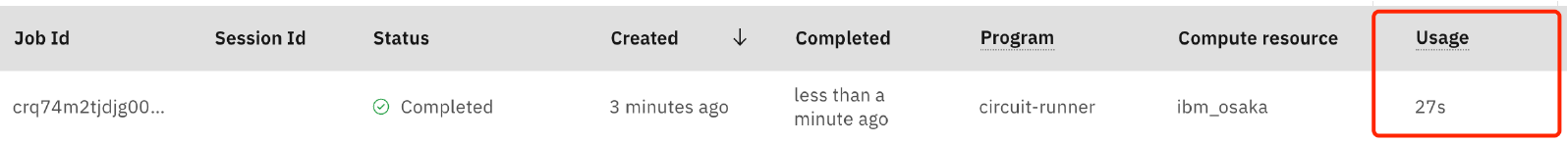}
        \caption{Actual Execution Time}
        \label{fig:actual}
    \end{subfigure}
    \caption{Estimated and Actual Execution Times on IBM Quantum Platform}
    \label{fig:Estimated and Actual Execution Times on IBM Quantum Platform}
\end{figure}

When a quantum circuit is submitted for execution on a quantum computing platform such as the IBM Quantum Platform, the platform usually provides an estimated execution time. The execution time includes loading the quantum circuit into the pre-selected quantum computer that creates the pulses, running multiple repetitions of calibration pulses and the quantum circuit, and resetting the qubits (relaxation) along with either calibration or executing the quantum circuit. These steps together make up the total execution time. As illustrated in Fig.~\ref{fig:estimated}, on the IBM quantum computer \texttt{ibm\_osaka}, the quantum circuit is expected to run for 2 minutes and 27.5 seconds. After the execution, the platform provides the actual execution time of the circuit. As shown in Fig.~\ref{fig:actual}, the actual execution time is only 27 seconds, substantially different from the estimation. The primary objective of our approach is to accurately predict the execution time, aiming to substantially reduce the significant errors observed in the estimates provided by quantum computing platforms. Please note that currently, IBM Quantum Platform only provides the worst case estimated execution time, which serves primarily as an upper bound for latency-critical scenarios. However, this estimate alone is insufficient for many practical use cases where developers and platform operators require more realistic predictions. Our method is expected to complement IBM's worst-case latency estimates by providing precise average-case execution times, which are essential for realistic cost estimation; developers can rely on IBM’s upper bounds when strict real-time guarantees are required and leverage our predictions for practical runtime planning. While the IBM Quantum Platform serves as a case study in this paper, our goal extends beyond this to a more general application across various platforms.

\section{Data Preparation}
\label{sec:data-preparation}
In this section, we describe our data preparation process, including the selection of the quantum circuits and quantum backends, as well as the measurement of execution time.

\subsection{Quantum circuit selection} \label{sec:prep:program}
We select our subject quantum circuits from the MQT Benchmark~\cite{Quetschlich2023mqtbench} which provides an extensive collection of quantum circuits covering various quantum computing applications across different levels of abstraction. This library comprises 3,000 quantum circuits, spanning from 2 to 130 qubits. Notably, this dataset exhibits significant diversity as it considers two qubit technologies, seven different backends, two compilers, and six corresponding settings. This diversity renders it a robust and versatile resource for our research that aims to estimate quantum execution time for diverse quantum circuits.
Due to the limitation of quantum computing resources, we must select a subset for our experiments. For the compiler, we have chosen Qiskit, an open-source SDK tailored for quantum computing. Qiskit specializes in circuit, pulse, and algorithm levels, enabling the creation and manipulation of quantum circuits. These circuits can be executed on prototype quantum backends available through the IBM Quantum Platform or via simulators on local machines. Adhering to the circuit model of universal quantum computation, Qiskit is versatile, supporting various quantum hardware types that comply with this model, including superconducting qubits and trapped ions.

As for the abstraction level selection (the different stages of compilation mentioned in Section~\ref{sec:background:compilation}), we choose the Target-independent Level. This is the compilation level for quantum circuits that end users create directly by writing code on Qiskit. For most users who lack extensive knowledge about quantum computers, there is no need for them to know how quantum circuits are compiled onto quantum computers. Therefore, selecting this compilation level is more meaningful, practical, and applicable to our research.

Finally, as for the quantum circuits, we selected all quantum circuits ranging from 2 qubits to 127 qubits (the limit of our considered quantum backends). When executing these quantum circuits, we found that a small portion encountered loading or compilation errors, and some had execution times that were excessively long, exceeding 10 minutes. Therefore, we excluded these quantum circuits with errors and set a timeout function to filter out those with excessively long execution times (over 10 minutes). Ultimately, we obtained a total of 1,510 subject quantum circuits.

\subsection{Quantum backend selection} \label{sec:prep:backend}
We consider the quantum backends provided by the IBM Quantum Platform as it is one of the mostly used quantum computing platforms and provides open accesses for free. Importantly, all the real devices and corresponding simulators used in this study are based on the superconducting qubit technology, while other hardware modalities (e.g., ion-trap processors) are outside the scope of our evaluation. For open accesses, the accessible quantum computers include \texttt{ibm\_brisbane}, \texttt{ibm\_osaka}, and \texttt{ibm\_kyoto} (at the time of writing), all of which support 127 physical qubits. Due to the high traffic on \texttt{ibm\_brisbane} and in order to ensure data diversity, we have chosen to execute the quantum circuits on both \texttt{ibm\_osaka} and \texttt{ibm\_kyoto}. To maintain consistency in data dimensions while leveraging simulation data and active learning to select samples for execution on real backends, we consider only simulators that support the same number of qubits as the real backends. Consequently, we selected simulators that support 127 qubits, namely \texttt{FakeWashington} and \texttt{FakeSherbrooke}. In summary, we consider the following backends:

\subsubsection{Quantum Computers}
\begin{itemize}
    \item \texttt{ibm\_osaka} (127 qubits)
\end{itemize}
\begin{itemize}
    \item \texttt{ibm\_kyoto} (127 qubits)
\end{itemize}
\subsubsection{Quantum Simulators}
\begin{itemize}
    \item \texttt{FakeWashington} (127 qubits)
\end{itemize}
\begin{itemize}
    \item \texttt{FakeSherbrooke} (127 qubits)
\end{itemize}

\subsection{Measurement of execution time}
In order to understand the execution time of quantum circuits, build and evaluate our approach for quantum execution time estimation, we execute our subject quantum circuits (see \ref{sec:prep:program}) on the selected backends (see \ref{sec:prep:backend}) and measure the execution time.

\noindent\textbf{Measurement of execution time on simulators.} 
When users want to run a quantum circuit on a simulator, they need to select a simulator (e.g., \texttt{FakeSherbrooke}) and compile the circuit with the simulator as the compilation target. Then, the compiled circuit should be submitted to the simulator for execution. The total time consumed in the process of execution is what we measure as the execution time of a quantum circuit on a simulator. Qiskit provides an API\footnote{https://docs.quantum.ibm.com/api/qiskit/0.45/execute} that makes it easy to obtain the time used by a simulator to execute a quantum circuit. Each quantum circuit is executed for 1024 shots on each simulator.

\noindent\textbf{Measurement of execution time on quantum computers.} 
Similarly, when users choose to execute a quantum circuit on a quantum computer, such as using the \texttt{ibm\_osaka} provided by the IBM Quantum Platform, they need to submit the designed and compiled quantum circuit to the IBM Quantum Platform, select \texttt{ibm\_osaka} as the quantum computer, and then queue for its execution. It is important to note that in order to select quantum computers provided by the IBM Quantum Platform to execute a quantum circuit, a token must be provided. Every IBM Quantum Platform account possesses this token, which grants access to the quantum computers on the platform. This token allows users to submit the designed and compiled quantum circuit to a specified quantum computer for execution.

Finally, the total time spent during the execution process is what we measure as the execution time of a quantum circuit on a quantum computer. The process of retrieving the execution time is similar to that on a simulator. We use the same API to obtain the execution time of quantum circuits on quantum computers as we do for simulators.

To ensure consistency in results, each quantum circuit is also executed for 1024 shots on quantum computers.

\subsection{Determining minimum repeated measures for quantum execution time accuracy}
In order to compensate for measurement errors, we ascertain the lowest requisite number of repeated measures for the execution time of each quantum circuit to maintain a satisfactory precision level, as delineated by Equation ~\ref{eq:Determining Minimum Repeated Measures} \cite{bookJainRaj}.

\begin{equation}
n = \left\lceil \left( \frac{100 \times z_{1-\alpha/2} \times s}{r \times \bar{x}} \right)^2 \right\rceil
\label{eq:Determining Minimum Repeated Measures}
\end{equation}

In Equation~\ref{eq:Determining Minimum Repeated Measures}, \( n \) signifies the required number of observations; \( z_{1-\alpha/2} \) corresponds to the critical value for the normal distribution at the chosen confidence interval \( (1 - \alpha) \); \( s \) denotes the standard deviation of the execution times obtained from each quantum circuit; \( r \) reflects the desired precision. The desired accuracy of \( r \) percent implies that the confidence interval should be \( (\bar{x}(1 - r/100), \bar{x}(1 + r/100)) \); and \( \bar{x} \) represents the mean of the execution times for each quantum circuit.

The value of \( r = 25 \) was set to balance the need for reasonable accuracy with the limitations of available computational resources, ensuring that the sample size remains manageable without excessively straining resources. With a confidence level of 95\%, the value of \( z \) is 1.960. Upon substituting \( \bar{x} \) and \( s \) for each quantum circuit, we ascertain that for over 90\% of quantum circuits, the value of \( n \) is \( \leq 3 \). Consequently, for each quantum circuit, we perform three executions on each simulator and quantum computer and then take the average to determine the execution time on that corresponding backend.

\section{Quantum execution time prediction model}
\label{sec:model}

In this section, we describe how we build and evaluate our quantum execution time prediction model, including the representation of the quantum circuits, the construction of the prediction model, the active learning for collecting training data for quantum circuits running on real quantum computers, as well our evaluation method and metrics.

\subsection{Quantum circuit representation} \label{sec:features}
To construct an ML model that captures the characteristics of quantum circuits, it is necessary to convert quantum circuits into what are known as feature vectors: vectors composed of integer and floating-point values, which render the quantum circuits compatible with effective model training. To effectively capture the characteristics of quantum circuits, we consider two categories of features: global features that capture the overall characteristics of a circuit and graph features that capture the graph structure of a circuit.
 
Below, we provide a detailed description of these two types of features.

\subsubsection{Global features}
\begin{figure} [h]
    \centering
    \includegraphics[width=1\linewidth]{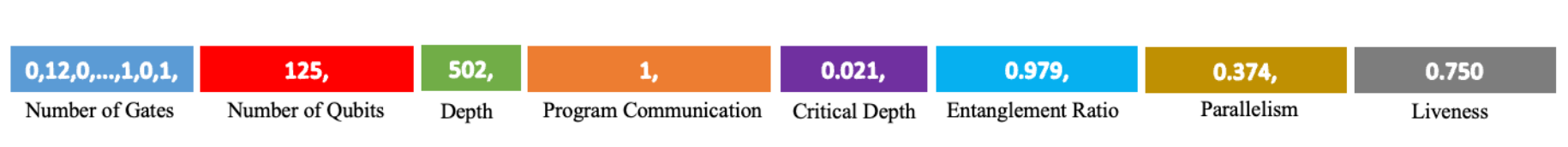}
    \caption{An example of the global features of a quantum circuit (Dimensionality: 41)}
    \label{fig:Global Features of A Quantum Circuit}
\end{figure}

Figure~\ref{fig:Global Features of A Quantum Circuit} shows an example of the global feature vector of a quantum circuit. 

Intuitively, for a quantum circuit, if it contains a large number of quantum gates, especially two-qubit gates (such as controlled-NOT gates), its execution time will be longer. Similarly, the number of qubits in the circuit can also affect execution time. Therefore, we take into account the number of different quantum gates; a gate is included if it appears in a quantum circuit, with its corresponding value being the absolute number of occurrences in that circuit, and zero if it does not appear. In our study, all quantum circuits include a total of 34 different quantum gates. Additionally, we consider the number of qubits, which is the total count of qubits in a quantum circuit, and the depth of the quantum circuit, which measures the longest path between the data input and output.

Additionally, based on previous related research, we also consider five other features to characterize the global attributes of each quantum circuit. These features include program communication~\cite{black2002quantum}, critical-depth, entanglement-ratio~\cite{GUZMAN2004382}, parallelism~\cite{murali2021instruction},
and liveness~\cite{viola1999dynamical}, as originally introduced in prior work~\cite{tomesh2022supermarq}. This methodical approach allows for a nuanced and detailed representation of the quantum circuits, facilitating more accurate and efficient training of the machine learning model.

\textbf{Program Communication} measures the amount of interaction required between qubits in a quantum circuit. It is quantified by calculating the average degree of interaction between qubits, where a qubit's "degree" refers to the number of other qubits it interacts with through multi-qubit operations~\cite{li2019tackling}. This measure helps assess how communication-intensive a quantum program is: sparsely connected applications will have values near zero, while denser programs will be close to one.

\textbf{Critical-Path} in a quantum circuit is the longest chain of dependent operations from start to finish. This path is crucial because it involves two-qubit operations, which are more error-prone and time-consuming~\cite{erata2024quantum, 9678792}. The critical depth, calculated as the ratio of two-qubit operations on this path to the total in the circuit, indicates how sequentially dependent the circuit is; a value close to 1 means the circuit is highly serialized.

\textbf{Entanglement-Ratio} reflects the extent of entanglement within a circuit. Since entanglement is fundamental to quantum computing, this ratio serves as a valuable indicator of performance. It is approximated by the ratio of two-qubit interactions to the total number of gate operations in the circuit.

\textbf{Parallelism} refers to the degree to which quantum operations can be performed simultaneously. Parallel operations can speed up execution~\cite{murali2021instruction}. The parallelism feature is determined by comparing the ratios of the number of qubits ($n$), the number of gates ($n_g$), and the circuit depth ($d$):
\begin{equation}
P = \left( \frac{n_g}{d} - 1 \right) \cdot \frac{1}{n - 1}
\end{equation}
This comparison helps assess how much parallelism is present in the circuit. Highly parallel circuits, which perform many operations in parallel, will have a parallelism feature value close to 1.

\textbf{Liveness} in a quantum circuit refers to the state of qubits during program execution. A qubit can either be actively involved in computation or remain idle, waiting for the next instruction. Ideally, a qubit should maintain its state while idling, but in reality, environmental interactions can cause the qubit to lose coherence ~\cite{viola1999dynamical}. The liveness feature captures how often qubits are actively used versus how often they are idle during the execution of a quantum circuit. This feature is calculated by forming a liveness matrix. In this matrix, each row represents a qubit, and each column corresponds to a time step in the circuit's execution. The matrix entries are 1 if the qubit is active and 0 if it is idle. To convert this matrix into a single scalar value representing liveness, we compute the average number of active qubit-time pairs. This is done by summing all entries in the liveness matrix and dividing by the total number of elements, which is the product of the number of qubits and the circuit depth. The value ranges from 0 to 1, where a higher value indicates more active usage of qubits during execution.

Quantum circuits with a higher proportion of two-qubit gates tend to have longer execution times. This is because two-qubit gates are generally more complex and time-consuming to execute than single-qubit gates. Additionally, a quantum circuit with longer average idle times might also experience longer execution times. If qubits are frequently idle, it might indicate inefficiencies in the circuit design, where operations could potentially be better parallelized or optimized to reduce waiting times.

Among the five features mentioned, program communication, critical depth, and entanglement ratio reflect the impact of the proportion and number of two-qubit gates on the execution time of quantum circuits from different perspectives. On the other hand, parallelism and liveness are likely related to the average idle times of qubits. We introduced these five features considering their potential influence on the execution time of quantum circuits.

Ultimately, the dimensionality of these features is 41, with the first 34 dimensions representing 34 different quantum gates. The remaining 7 dimensions correspond to the following features: number of qubits, depth, program communication, critical depth, entanglement ratio, parallelism, and liveness.

\subsubsection{Graph features}
A quantum circuit can be conceptualized as a graph, allowing for the extraction of graph-based information for analytical purposes. The Graph Transformer \cite{velickovic2017graph, yun2019graph}, derived from attention-based \cite{vaswani2017attention} Transformer models renowned for their effectiveness in sequence modeling, represents a robust method for this type of analysis. By leveraging the attention mechanism, the Graph Transformer updates the features of each node in successive layers. This approach has been applied in recent studies to estimate the fidelity of quantum circuits, where each node in the graph represents a qubit, quantum gate, or measurement, and edges denote the time-dependent sequence of operations \cite{wang2022quest}. This graph-based representation facilitates a nuanced understanding of the circuit's dynamics and interactions.

\begin{figure} [h]
    \centering
    \includegraphics[width=0.64\linewidth]{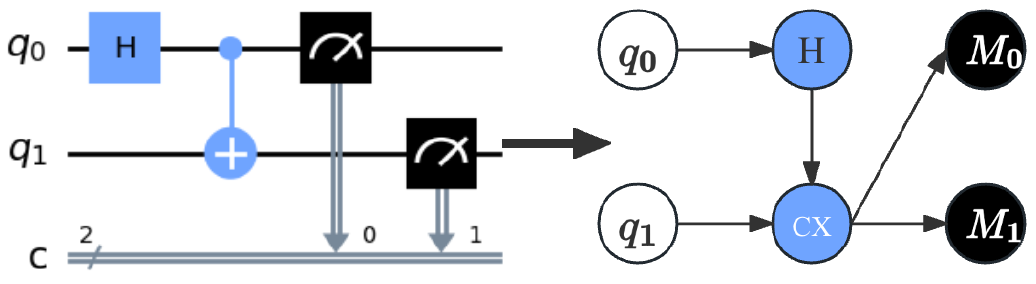}
    \caption{An example of converting a quantum circuit into a directed acyclic graph (dag)}
    \label{fig:dag}
\end{figure}

We initially model the structure of quantum circuits using directed acyclic graphs (DAGs), where each node signifies either a qubit, a quantum gate or a measurement, and edges delineate the sequential dependencies among qubits and gates. This structure is encapsulated within an adjacency matrix that captures the circuit's connectivity. An example of converting the quantum circuit in Fig.~\ref{fig:An example of a Quantum Circuit} into a DAG is shown in Fig.~\ref{fig:dag}. The node labeled with CX in the middle column represents a CNOT gate, while the last column's \(M_0\) indicates a measurement on qubit \(q_0\), and \(M_1\) indicates a measurement on qubit \(q_1\).

\begin{figure} [h]
    \centering
    \includegraphics[width=0.65\linewidth]{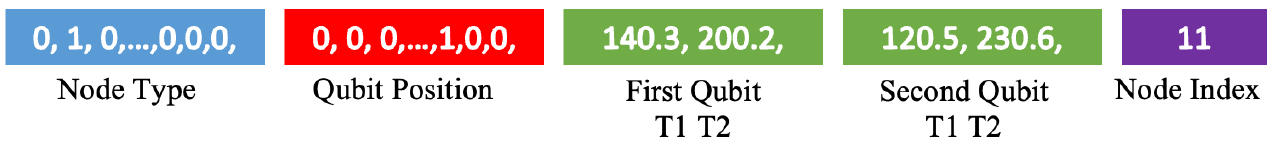}
    \caption{An example node of the graph features of a quantum circuit (Dimensionality: 178)}
    \label{fig:Graph Features of A Quantum Circuit}
\end{figure}

For every node within the DAG, we assign a feature vector encapsulating attributes such as the node's type, the qubit's location, decay constants T1 and T2 for the concerned qubit(s) (a gate may involve one or two qubits), and the gate's sequential position, as delineated in Fig.~\ref{fig:Graph Features of A Quantum Circuit}.

In our experimentation, we align the maximum number of qubits to 127, mirroring the upper limit supported by contemporary quantum hardware, resulting in feature vectors spanning 178 elements. The leading 46 entries constitute a one-hot encoding for the node classification, including the initial qubit state, measurements, and 44 distinct gate types. Subsequently, 127 entries demarcate the qubit(s)' positions, with a "1" denoting the qubit's involvement in a gate operation. This scheme extends to gates influencing multiple qubits. Note that the quantum circuits used in this experiment include only single- and two-qubit gates. Calibration details for the backend are represented by four entries formatted as [T1, T2 for the primary target qubit, T1, T2 for the secondary target qubit]. Inapplicable features for specific nodes are assigned a zero value. For instance, a gate like X, which impacts only one qubit, would have the T1 and T2 values for a second qubit set to zero. The final entry of the vector is reserved for the node's index. Fig.~\ref{fig:Graph Features of A Quantum Circuit} illustrates the complete composition of the feature vector.

Finally, we input both parts of the feature set into the graph transformer model for training and prediction.

\subsection{Model construction for execution time prediction} \label{sec:model-construction}

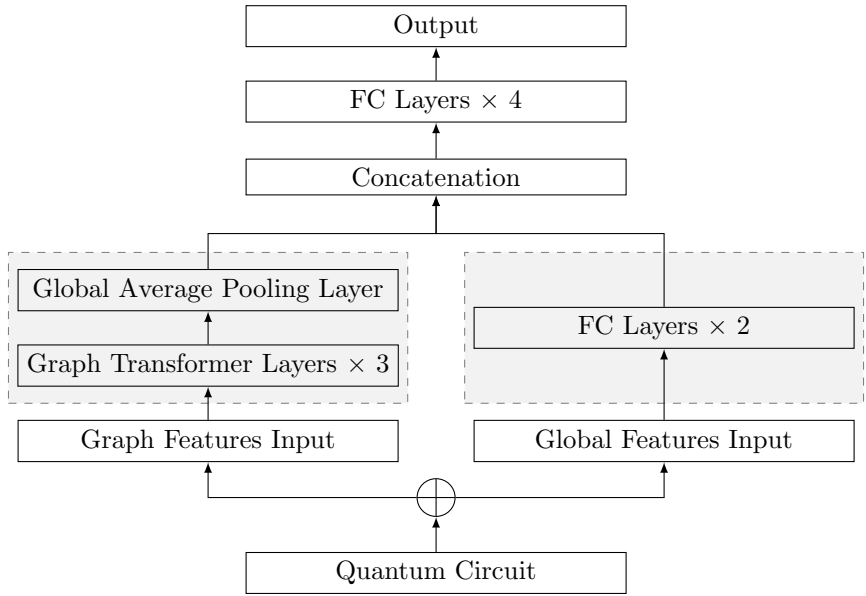
\begin{figure}[ht]
\centering
\begin{tikzpicture}
[
    block/.style={rectangle, draw, minimum width=5cm, minimum height=0.4cm},
    line/.style={draw, -latex},
    background/.style={rectangle, draw=black!50, dashed, fill=gray!10, minimum width=1cm,   % 设置最小宽度
        minimum height=2cm}, % 增加了填充色
    circle node/.style={circle, draw, inner sep=0pt, minimum size=0.5cm, 
    path picture={
            \draw 
            (path picture bounding box.south) -- (path picture bounding box.north)
            (path picture bounding box.west) -- (path picture bounding box.east);
        }},
]

% quantum circuit
\node[block] (qc) at (3,-1.75) {Quantum Circuit};

% graph features
\node[block] (gf1) at (0,0) {Graph Features Input};
\node[block] (gf2) at (0,1) {Graph Transformer Layers $\times$ 3};
\node[block,] (ga) at (0,2) {Global Average Pooling Layer};

\draw[line] (gf1) -- (gf2);
\draw[line] (gf2) -- (ga);

% global features
\node[block] (gf3) at (6,0) {Global Features Input};
\node[block] (fc1) at (6,1.5) {FC Layers $\times$ 2};

\draw[line] (gf3) -- (fc1);

% Add circle nodes with crosses
\node[circle node] (c1) at (3,-0.75) {};
\draw[line] (qc) -- (c1);
\draw[line] (c1) -- (0,-0.75) --++ (gf1);
\draw[line] (c1) -- (6,-0.75) --++ (gf3);

% Concatenation
\node[block] (co) at (3,3.5) {Concatenation};
\node[block] (fc2) at (3,4.5) {FC Layers $\times$ 4};
\node[block] (ou) at (3,5.5) {Output};

\draw[line] (ga) -- (0,2.75) -- (3,2.75) -- (co);
\draw[line] (fc1) -- (6,2.75) -- (3,2.75) -- (co);
\draw[line] (co) -- (fc2);
\draw[line] (fc2) -- (ou);

% Background shapes
\begin{scope}[on background layer]
    \node[background, fit=(gf2) (ga)] {};
    \node[background, fit=(fc1)] {};
\end{scope}
\end{tikzpicture}
\caption{The neural network architecture of the model}
\label{fig:The neural network architecture of the model}
\end{figure}

Fig.~\ref{fig:The neural network architecture of the model} shows the neural network architecture of the model we developed to predict the execution time of quantum circuits. For a quantum circuit, the dimension of its graph features is \( n \times 178 \), where \( n \) represents the number of nodes, and each quantum circuit has a different number of nodes. Each set of graph features passes through three layers of Graph Transformer, and then through Global Average Pooling to reduce the dimensions to \( 1 \times 178 \). On the other hand, the dimension of a quantum circuit's global features is \( 1 \times 41 \). The global features first pass through two fully connected layers, transforming their dimensions to \( 1 \times 64 \). These are then concatenated with the output from the Global Average Pooling. The dimensions of the concatenated results are \( 1 \times 242 \). This combined data is then fed into four additional fully connected layers to produce the final output, which is the predicted execution time of a quantum circuit on a backend, either a simulator or a quantum computer.
When training the model, we utilized grid search to determine the optimal hyperparameters. 
Specifically, we split the dataset into training, validation, and test sets in an 8:1:1 ratio, with each partition kept fixed throughout the experiments. The training set is used to train the model, while the validation set is used to select the optimal hyperparameters. We vary the number of epochs from 100 to 800 in increments of 100, and the batch size from 32 to 160 in increments of 32. For each combination of hyperparameters, the model is retrained from scratch to avoid any data leakage. After evaluating performance across all combinations, we find that setting the number of epochs to 500 and the batch size to 128 produces the best results.

\subsection{Active learning for quantum execution time prediction} \label{sec:active-learning}

\subsubsection{Active learning and greedy sampling techniques}

Active learning, a branch of machine learning, enables algorithms to selectively choose their training data, which is particularly useful when labels are costly or hard to obtain, in our case, collecting the execution time of quantum circuits. This method enhances learning efficiency by maximizing accuracy with fewer labeled instances, thus reducing the overall cost of data labeling. It is especially relevant in scenarios where there is ample data but few labels \cite{settles2009active}. Within active learning frameworks, Greedy Sampling (GS) plays a pivotal role in refining machine learning models.

Greedy Sampling is a foundational technique where a model strategically chooses new samples to enhance accuracy rapidly with a minimal number of labeled instances, focusing on metrics such as informativeness, representativeness, and diversity to enrich the training dataset \cite{wu2019active}.
A brief summary of greedy sampling techniques is as follows:

\begin{itemize}
   \item \textbf{GS on the input domain (\(G_x\))}: This approach increases diversity in the input space by first selecting samples nearest to the centroid of the unlabeled pool and then incrementally adding those furthest away, fostering a diverse training set \cite{wu2019active}.
  
  \item \textbf{GS on the output domain (\(G_y\))}: Concentrating on the output domain, \(G_y\) aims to diversify the range of output values. It utilizes the predicted outputs from a regression model to gauge and incorporate diversity in the dataset \cite{wu2019active}.
  
  \item \textbf{GS on input and output (\(G_{xy}\))}: Integrating the strategies of \(G_x\) and \(G_y\), \(G_{xy}\) augments diversity across both input and output spaces using a combined distance metric to optimize the selection of samples \cite{wu2019active}.
\end{itemize}

Due to the limited quantum computing resources (e.g., 10 minutes per month of free access to IBM Quantum's computers) and the lengthy queue time (e.g., several hours) before each quantum circuit can be executed, it is impractical to run all quantum circuit experiments directly on these quantum computers. Therefore, we employ active learning strategies to perform random sampling of the quantum circuits. This method allows for efficient utilization of available quantum computing resources while ensuring robust sampling coverage.

\subsubsection{Active learning for collecting execution time of quantum circuits} \label{sec:active-learning}
As described in Section~\ref{sec:data-preparation}, we selected 1510 quantum circuits and two quantum computers (\texttt{ibm\_osaka} and \texttt{ibm\_kyoto}), constituting a dataset of 3020 (\(1510 \times 2\)) data points. At a confidence level of 95\% and an error margin of 5\%, we need to sample 340 out of the 3020 available. For greedy sampling, we opt for the \( G_x \) method, which iteratively selects samples based on the features of the input samples until the required sample size is achieved.

For the features of the input samples, namely the quantum circuits, as described earlier (see \ref{sec:features}), we divide them into two categories: global features and graph features.
We first apply min-max scaling to normalize the two feature sets, rescaling their ranges to [0, 1] to prevent varying scales from interfering with the results. Then, we assign an equal weight of 0.5 to each set.
Consequently, we start by identifying the sample closest to the center of the sample space in Euclidean distance as our initial sample. Subsequently, in each iteration, we select the sample that is furthest in Euclidean distance from the already chosen set of samples. The selected sample is then updated into the sample set, and this process continues until 340 samples are selected.

Consider a pool consisting of $N$ samples, specifically quantum circuits, denoted by $\{x_n\}_{n=1}^N$. Initially, all samples are unlabeled. The primary objective is to select and label a subset of $K$ samples from this ensemble. The selection process, governed by \(G_x\), commences by identifying the initial sample for labeling as the one closest to the centroid of all $N$ samples, thus minimizing the Euclidean distance to the centroid, which is calculated as the geometric mean of the sample coordinates.

Assuming, without loss of generality, that the first $k$ samples have already been selected, the selection mechanism \(G_x\) proceeds to evaluate each of the remaining $N-k$ unlabeled samples, denoted as $\{x_n\}_{n=k+1}^N$. For each of these samples, \(G_x\) computes the distance to each of the already labeled $k$ samples. This iterative selection process is designed to incrementally choose the remaining $K-1$ samples. For each of the remaining $N - k$ unlabeled samples $\{x_n\}_{n=k+1}^N$, \(G_x\) computes first its distance to each of the $k$ labeled samples:

\begin{equation}
\label{eq:distance}
\begin{split}
d_{x_{nm}} = \|x_n - x_m\|= \alpha \dot \|x_{n_{global}} - x_{m_{global}}\| +(1 - \alpha) \dot \|x_{n_{graph}} - x_{m_{graph}}\|,\\
\quad m = 1, \ldots, k; \quad n = k + 1, \ldots, N; \quad \alpha \in [0, 1]
\end{split}
\end{equation}
where $x_{n_{\text{global}}}$ represents the normalized global features of the $n$-th quantum circuit, and $x_{n_{\text{graph}}}$ denotes the normalized graph features of the same circuit, $\alpha$ is a parameter that lies within the interval [0, 1]. For the purposes of this paper, $\alpha$ is set at 0.5 for a balanced weight for the two feature dimensions.
Then we obtain $d_{x_n}$, the shortest distance from $x_n$ to all $k$ labeled samples:
\begin{equation} 
\label{eq:min-dist}
d_{x_n} = \min_{m} d_{x_{nm}}, \quad n = k + 1, \ldots, N
\end{equation}
and finally selects the sample with the maximum $d_{x_n}$ to label. The pseudo-code is given in Algorithm ~\ref{alg:alg1}.

\begin{algorithm}
\caption{The GSx Active Learning method, modified from GS in \cite{wu2019active, yun2019graph}.}
\label{alg:alg1}
\begin{algorithmic}
\State \textbf{Input:} $N$ unlabeled samples: $Z = \{x_n\}_{n=1}^N$;
$K$, the maximum number of labels to query.
\State \textbf{Output:} $K$ labeled samples: $S$.

\State 
\State // Initialize the first selection
\State Set $Z = \{x_n\}_{n=1}^N$ and $S = \emptyset$;
\State Identify $x'$, the sample closest to the centroid of $Z$;
\State Move $x'$ from $Z$ to $S$;
\State Re-index the sample in $S$ as $x_1$, and the samples in $Z$ as $\{x_n\}_{n=2}^N$;
\State 
\State // Select $K-1$ more samples incrementally
\For{$k = 1, \ldots, K-1$}
    \For{$n = k+1, \ldots, N$}
        \State Compute $d_{x_n}$ according to Equations~\eqref{eq:distance} and \eqref{eq:min-dist};
    \EndFor
    \State Identify the $x'$ that has the largest $d_{x_n}$;
    \State Move $x'$ from $Z$ to $S$;
    \State Re-index the samples in $S$ as $\{x_m\}_{m=1}^{k+1}$, and the samples in $Z$ as $\{x_n\}_{n=k+2}^N$;
\EndFor
\State 
\State Query to label all $K$ samples in $S$.
\end{algorithmic}
\end{algorithm}

For the same quantum circuit compiled for different quantum computers, the graph structure varies, altering graph features while global features remain constant. Considering both feature types during greedy sampling, we have 3020 data points for 1510 quantum circuits. During selection, a circuit may be chosen twice, corresponding to compilations for \texttt{ibm\_osaka} and \texttt{ibm\_kyoto}. Of the 340 sampled circuits, 192 need to be executed on \texttt{ibm\_osaka} and 148 on \texttt{ibm\_kyoto}, with each circuit run three times to average execution times.

\subsection{Evaluation} \label{sec:evaluatino}

\subsubsection{Evaluation method}
We begin by training the model on a dataset comprising execution times from 1510 quantum circuits across two simulators, yielding 3020 data points. This dataset is divided into training and test sets in a 9:1 ratio. We then select and save the best-performing model for the test set, characterized by the lowest MSE (Mean Squared Error). To minimize the use of costly real quantum hardware, this model serves as a pre-trained model and is subsequently fine-tuned using a small set of execution time data collected from real quantum computers. Next, we employ the saved pre-trained model to conduct 10-fold cross-validation on a dataset of 340 data points, chosen through greedy sampling from the dataset of execution times collected from quantum circuits executed on quantum computers. This process begins by randomly shuffling the dataset and dividing it into ten equal folds. For each fold, one part is designated as the test set, while the remaining nine folds (90\% of the data) are further split into a training subset (80\%) and a validation subset (10\%) using a 8:1 ratio. We fix the hyperparameters to a batch size of 32 and 500 training epochs. After each training epoch, we evaluate the latest model on the validation subset (we obtain a new model after each epoch); finally, the best-performing model (i.e., the one with the lowest MSE on the validation subset) is selected. This selected model is then evaluated on the independent test set. The average of these results across the ten test folds is reported as the final performance metric for predicting the execution times of quantum circuits on quantum computers.

\subsubsection{Evaluation metrics} \label{sec:eval-metrics}
We have selected R-squared, MSE (Mean Squared Error), and NMSE (Normalized Mean Squared Error) to evaluate the performance of the model we obtained.

\noindent\textbf{R-squared (\( R^2 \))}, or the coefficient of determination, is a statistical measure used to evaluate the goodness of fit of a regression model. It indicates the proportion of variance in the dependent variable that is predictable from the independent variables. \( R^2 \) can be expressed as:

\begin{equation}
    R^2 = 1 - \frac{\sum (y_i - \hat{y}_i)^2}{\sum (y_i - \bar{y})^2}
\end{equation}

\noindent where \( y_i \) represents the actual execution time of quantum circuit \( i \), while \( \hat{y}_i \) represents the predicted execution time for quantum circuit \( i \) as estimated by our model. Additionally, \( \bar{y} \) is the mean of the actual execution times for all quantum circuits. 

Normally\footnote{Unfortunately, not always, as explained soon.}, \( R^2 \) values range from 0 to 1.
A result closer to 1 indicates that our model can accurately estimate the execution time of quantum circuits on a specific backend. Conversely, a result closer to 0 suggests that our model has poor predictive ability.
However, \( R^2 \) value can also be negative. This occurs when the predictions are worse than a simple model that predicts using the mean of the dependent variable. A negative \( R^2 \) value indicates very poor predictive accuracy, suggesting that the model fails to capture the essential trends of the data and is unsuitable for making predictions.

\noindent\textbf{Mean Squared Error (MSE)} is a widely used measure of the average squared difference between the estimated values and the actually observed values.
MSE is given by:
\begin{equation}
    \text{MSE} = \frac{1}{n} \sum_{i=1}^{n} (y_i - \hat{y}_i)^2
\end{equation}

\noindent where \( y_i \) denotes the actual execution time of the quantum circuit indexed by \( i \), while \( \hat{y}_i \) denotes the model-estimated execution time for the same circuit.
A lower MSE value signifies a model that more accurately predicts the real execution times.

\noindent\textbf{Normalized Mean Squared Error (NMSE).}
Given that execution times for quantum circuits on quantum computers are typically short, even a model that is not exceptionally accurate might still produce a relatively small MSE value. Thus, we also measure the Normalized Mean Squared Error (NMSE) to provide a more accurate assessment. NMSE adjusts the mean squared error for the variance of the data, rendering it unitless and more interpretable across various datasets and scales.
NMSE is given by:
\begin{equation}
    \text{NMSE} = \frac{\text{MSE}}{\frac{1}{n} \sum_{i=1}^{n} (y_i - \bar{y})^2}
\end{equation}

\noindent where \( \bar{y} \) represents the average of the actual execution times of the quantum circuits and \( y_i \) the predicted execution time of the circuit \( i \).
A lower NMSE value suggests that the model's errors are minor relative to the data's variance.

\smallskip
\noindent\textit{Note.} While it is true that NMSE and \( R^2 \) are mathematically related by \( R^2 = 1 - \text{NMSE} \) under certain conditions, we intentionally include both metrics to improve readability and interpretability for different audiences. NMSE offers a normalized error scale that is directly comparable across datasets, while \( R^2 \) provides a goodness-of-fit measure commonly used in regression tasks. Reporting both helps highlight complementary perspectives on model performance.
\section{Results}
\label{sec:result}
This paper aims to understand and accurately predict the execution times of quantum circuits on platforms such as the IBM Quantum Platform, through answering three research questions. In this section, we present our approaches and discuss the results for each research question.

\subsection{RQ1: How do the execution times of quantum circuits vary across different circuits and backends?}

\subsubsection{Statistical analysis of quantum circuit execution times on simulators} \label{sec:simulator-time}

\begin{figure} [h]
    \centering
    \includegraphics[width=1\linewidth]{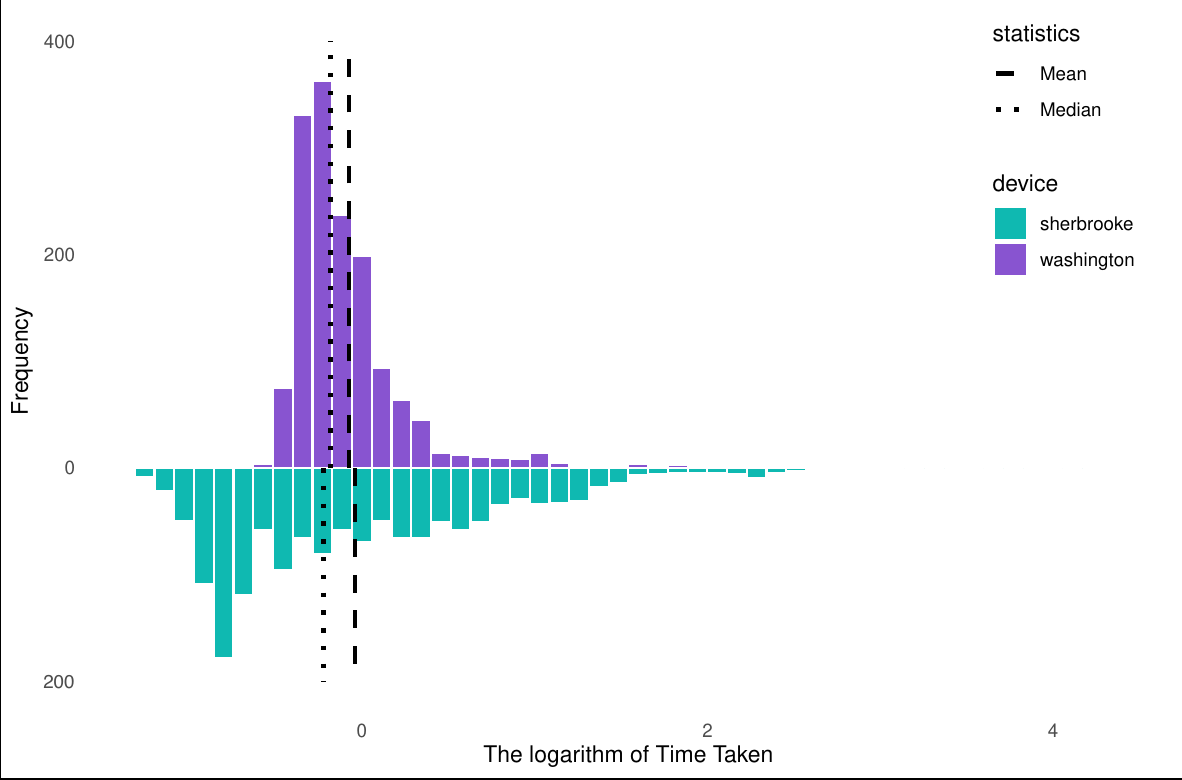}
    \caption{Histograms of execution time (in seconds) statistics for quantum circuits on different simulators}
    \label{fig:combined simulators}
\end{figure}

We executed the 1,510 selected quantum circuits on the two considered simulators (\texttt{FakeSherbrooke} and \texttt{FakeWashington}) and measured their execution times (see Section~\ref{sec:data-preparation}). 
Table~\ref{tab:execution time on simulators} and Figure~\ref{fig:combined simulators} show the distributions of the execution times on the two simulators. Table~\ref{tab:execution time on simulators} presents the summary statistics of the execution times of the 1,510 quantum circuits. Figure~\ref{fig:combined simulators} shows the histograms for these quantum circuits's execution times. Considering that the execution time values are skewed, we use the logarithm of the execution times to facilitate the observation of their distributions. The reason for the apparent blank space on the right side of Figure~\ref{fig:combined simulators} is due to the presence of individual quantum circuits whose execution times on simulators are exceptionally high, with the maximum value being 73.24 seconds. After taking the natural logarithm, this value is approximately 4.294.

\noindent\textbf{The execution times of quantum circuits on simulators vary significantly (e.g., up to 271 times difference) across different circuits.} 
From Table~\ref{tab:execution time on simulators} and Figure~\ref{fig:combined simulators}, we observe that on both simulators, the execution times vary significantly across different circuits.
The execution times of the majority of the circuits are relatively short, with around half of the quantum circuits have execution times shorter than 1 second. 
From Fig.~\ref{fig:combined simulators}, we can observe that many values are negative, indicating execution times of less than one second.
Nonetheless, there are some quantum circuits with relatively long execution times. For example, on \texttt{FakeSherbrooke}, the longest execution time for a single quantum circuit is 73.24 seconds, while on \texttt{FakeWashington}, the maximum execution time for a single quantum circuit is 47.64 seconds. On \texttt{FakeSherbrooke} and \texttt{FakeWashington}, the execution times of the longest-running quantum circuits are approximately 271 and 86 times those of the shortest-running quantum circuits, respectively.
As shown in Fig.~\ref{fig:combined simulators}, the distribution of execution times of the quantum circuits is more concentrated on \texttt{FakeSherbrooke} than on \texttt{FakeWashington}.

\begin{table}[h]
\centering
\caption{Statistical information on quantum circuit execution times on simulators (in seconds)}
\label{tab:execution time on simulators}
\begin{tabular}{@{}lcc@{}}
\toprule
\textbf{Statistic} & \textbf{FakeSherbrooke} & \textbf{FakeWashington} \\ \midrule
Count & 1510 & 1510 \\
Mean & 1.50 & 1.12 \\
Standard Deviation & 2.72 & 2.05 \\
Minimum & 0.27 & 0.55 \\
25\% & 0.48 & 0.74 \\
50\% & 0.80 & 0.83 \\
75\% & 1.62 & 1.02 \\
Maximum & 73.24 & 47.64 \\ \bottomrule
\end{tabular}
\end{table}

\noindent\textbf{The execution times of the same circuits on different simulators are statistically different but highly correlated.}
We perform statistical tests to compare and correlate the execution times of the circuits on the two simulators.
Firstly, at a significance level of 0.05, we use the Shapiro-Wilk test \cite{shapiro1965analysis} to determine if the execution times of quantum circuits on \texttt{FakeSherbrooke} and \texttt{FakeWashington} each follow a normal distribution. The Shapiro-Wilk test is a statistical method used to assess if a dataset is normally distributed. It is widely used to check the normality assumptions in data before applying certain statistical tests.
The results of the Shapiro-Wilk Test are presented in the first two rows of Table~\ref{tab:satistical test results on different simulators}, with $p$-values significantly less than 0.05. Consequently, it can be concluded that the execution times of quantum circuits are not normally distributed on both \texttt{FakeSherbrooke} and \texttt{FakeWashington}.

Given the non-normal distributions, we then use the Wilcoxon Signed-Rank Test~\cite{Neuhäuser2011} to perform paired comparisons of the distributions of execution times of the same quantum circuit on the two different simulators. The Wilcoxon Signed-Rank Test is a nonparametric statistical test used to compare two related samples or matched pairs to determine whether the median difference between the pairs of observations is significantly different from zero.
As shown in the third row of Table~\ref{tab:satistical test results on different simulators}, the $p$-value is significantly less than 0.05. Hence, we conclude that there is a statistically significant difference between the execution times of the same quantum circuits on the two simulators.

Finally, we employ Spearman's Rank Correlation Coefficient measurement \cite{spearman1987proof} to examine whether there is a correlation in the execution times of quantum circuits on the two different simulators. Specifically, we investigate whether the same quantum circuit tends to have either consistently higher or consistently lower execution times on both simulators.
Spearman's Rank Correlation Coefficient measures the strength and direction of a relationship between two variables using their ranks. It is useful when the data is not normally distributed or when dealing with ordinal variables. The result, a coefficient (i.e., the $\rho$ value) ranging from -1 to 1, shows whether the relationship is positive, negative, or non-existent.
The results of the Spearman's Rank Correlation Coefficient measurement are shown in the fourth row of Table~\ref{tab:satistical test results on different simulators}. The $p$-value is significantly less than 0.05, leading us to conclude that there is a statistically significant correlation between the distributions. Moreover, with a $\rho$ value of 0.914, it is concluded that there is a strong positive correlation in the execution times of quantum circuits on the two simulators. Generally, this means that for the same quantum circuit, the execution times are consistently either both high or both low on the two simulators.

\begin{table}[h]
\centering
\caption{Statistical test results for comparing the circuits' execution times on different simulators}
\label{tab:satistical test results on different simulators}
\begin{tabular}{l|c}
\hline
\textbf{Test} & \textbf{Results} \\ \hline
Shapiro-Wilk Test on \texttt{FakeSherbrooke} & $p$-value $<2.2 \times 10^{-16}$ \\
Shapiro-Wilk Test on \texttt{FakeWashington} & $p$-value $<2.2 \times 10^{-16}$ \\
Wilcoxon Signed-Rank Test & $p$-value = $1.179 \times 10^{-12}$ \\
Spearman's Rank Correlation Coefficient & $p$-value $<2.2 \times 10^{-16}$; $\rho = 0.914$ \\
\hline
\end{tabular}
\end{table}

To investigate why the same quantum circuit yields different execution times on different simulators, we selected the circuit with the largest execution time discrepancy between the two simulators, namely \texttt{qwalk-noancilla\_indep\_qiskit\_9}, for further analysis. As shown in Table~\ref{tab:compilation_results}, we present the compilation results of this circuit on both \texttt{FakeSherbrooke} and \texttt{FakeWashington}. The results reveal that the number of ECR (Echoed Cross-Resonance) gates in the \texttt{FakeSherbrooke} compilation is comparable to the number of Controlled-Not (CX) gates in the \texttt{FakeWashington} compilation. Since both ECR and CX are two-qubit gates, this part of the circuit complexity remains similar across the two backends.
However, for other gate types—namely RZ, X, and SX—the circuit compiled for \texttt{FakeSherbrooke} contains a significantly higher number of gates compared to the compilation for \texttt{FakeWashington}. All gate-duration values are taken verbatim from each backend’s calibration snapshot\footnote{https://quantum.cloud.ibm.com/docs/en/api/qiskit/0.44/qiskit.providers.models.BackendProperties}. IBM realises an RZ gate as a virtual frame change, so its physical duration is effectively zero\footnote{https://quantum.cloud.ibm.com/docs/en/api/qiskit/0.44/qiskit.circuit.library.RZGate}. Moreover, single-qubit X-family gates (X, SX) share the same 160\,dt DRAG pulse envelope, giving them identical calibrated durations~\cite{erata2024quantum}. According to Table~\ref{tab:avg_gate_times}, both CX and ECR gates have similar average execution times, and their execution times are the longest among all gate types, which explains why the total execution times are high on both simulators. Nevertheless, due to the overwhelming number of SX and X gates introduced during the compilation for \texttt{FakeSherbrooke}, the overall execution time on \texttt{FakeSherbrooke} is significantly higher than that on \texttt{FakeWashington}.
\begin{table}[h]
    \centering
    \caption{Compilation results of quantum circuit \texttt{qwalk-noancilla\_indep\_qiskit\_9} on two simulators}
    \label{tab:compilation_results}
    \begin{tabular}{lcc}
        \toprule
        \textbf{Metric} & \textbf{\texttt{FakeWashington}} & \textbf{\texttt{FakeSherbrooke}} \\
        \midrule
        Execution Time (s) & 42.43 & 73.24 \\
        \midrule
        Gate Count: cx       & 18519 & --- \\ 
        Gate Count: ecr      & ---   & 18273 \\ 
        Gate Count: rz       & 9099  & 48126 \\
        Gate Count: sx       & 125   & 36291 \\
        Gate Count: x        & 10    & 4262 \\
        Circuit Depth & 19488 & 54509 \\
        \bottomrule
    \end{tabular}
\end{table}
\begin{table}[h]
    \centering
    \caption{Average gate execution times on two simulators}
    \label{tab:avg_gate_times}
    \begin{tabular}{lccccc}
        \toprule
        \textbf{Simulator} & \textbf{cx (ns)} & \textbf{ecr (ns)} & \textbf{rz (ns)} & \textbf{sx (ns)} & \textbf{x (ns)} \\
        \midrule
        \texttt{FakeWashington} & 550.41 & ---               & 0.0  & 35.56 & 35.56 \\
        \texttt{FakeSherbrooke} & ---    & 533.83           & 0.0  & 56.89 & 56.89 \\
        \bottomrule
    \end{tabular}
\end{table}

We further quantify how target-independent features relate to execution time on simulators. Specifically, we employ Spearman’s rank correlation to examine, on the two simulators, whether execution time is associated with the target independent features, namely \emph{depth}, \emph{num\_qubits}, \emph{two-qubit gate count}, and \emph{num\_gates}.
In addition, we use Pearson’s correlation coefficient~\cite{pearson1896vii} to assess whether these associations are linear. As shown in Table~\ref{tab:corr-sim}, \emph{depth} is moderately associated both monotonically and linearly: \texttt{FakeSherbrooke} ($\rho=0.730$, $r=0.636$) and \texttt{FakeWashington} ($\rho=0.677$, $r=0.306$). \emph{Num\_qubits} exhibits only modest monotonic and negligible linear association: \texttt{FakeSherbrooke} ($\rho=0.382$, $r=0.081$) and \texttt{FakeWashington} ($\rho=0.352$, $r=-0.082$). \emph{Two-qubit gate count} shows a monotonic but limited linear association with time: on \texttt{FakeSherbrooke} (Spearman $\rho=0.796$, Pearson $r=0.315$) and on \texttt{FakeWashington} ($\rho=0.722$, $r=0.046$). \emph{Num\_gates} behaves similarly to \emph{two-qubit gate count}: \texttt{FakeSherbrooke} ($\rho=0.795$, $r=0.355$) and \texttt{FakeWashington} ($\rho=0.722$, $r=0.062$). The weak linear relations of individual features indicate that single features are insufficient as predictors, so we need models that combine features. Moreover, the coefficients differ across backends; for example, on \texttt{FakeWashington} Pearson's $r$ is generally smaller (e.g., 0.046 vs.\ 0.315 for the \emph{two-qubit gate} feature) than on \texttt{FakeSherbrooke}, which suggests the need for backend dependent features and models.

\begin{table}[h]
\centering
\setlength{\tabcolsep}{3pt}
\caption{Monotonic vs.\ linear associations with execution time on simulators}
\label{tab:corr-sim}
\begin{tabular}{lrrrr}
\toprule
& \multicolumn{2}{c} {Spearman $\rho$}& \multicolumn{2}{c} {Pearson $r$}\\
\cmidrule(lr){2-3}\cmidrule(lr){4-5}
Feature & \textbf{\texttt{FakeSherbrooke}} & \textbf{\texttt{FakeWashington}} & \textbf{\texttt{FakeSherbrooke}} & \textbf{\texttt{FakeWashington}} \\
\midrule
Depth            & 0.730 & 0.677 & 0.636 & 0.306                  \\
Num\_qubits           & 0.382 & 0.352 & 0.081 & $-0.082$                \\
Two-qubit gate count & 0.796 & 0.722 & 0.315 & 0.046 \\
Num\_gates           & 0.795 & 0.722 & 0.355 & 0.062                \\
\bottomrule
\end{tabular}
\vspace{2pt}
\end{table}

\subsubsection{Statistical analysis of quantum circuit execution times on quantum computers}\label{sec:computer-time}
\begin{figure} [h]
    \centering
    \includegraphics[width=1\linewidth]{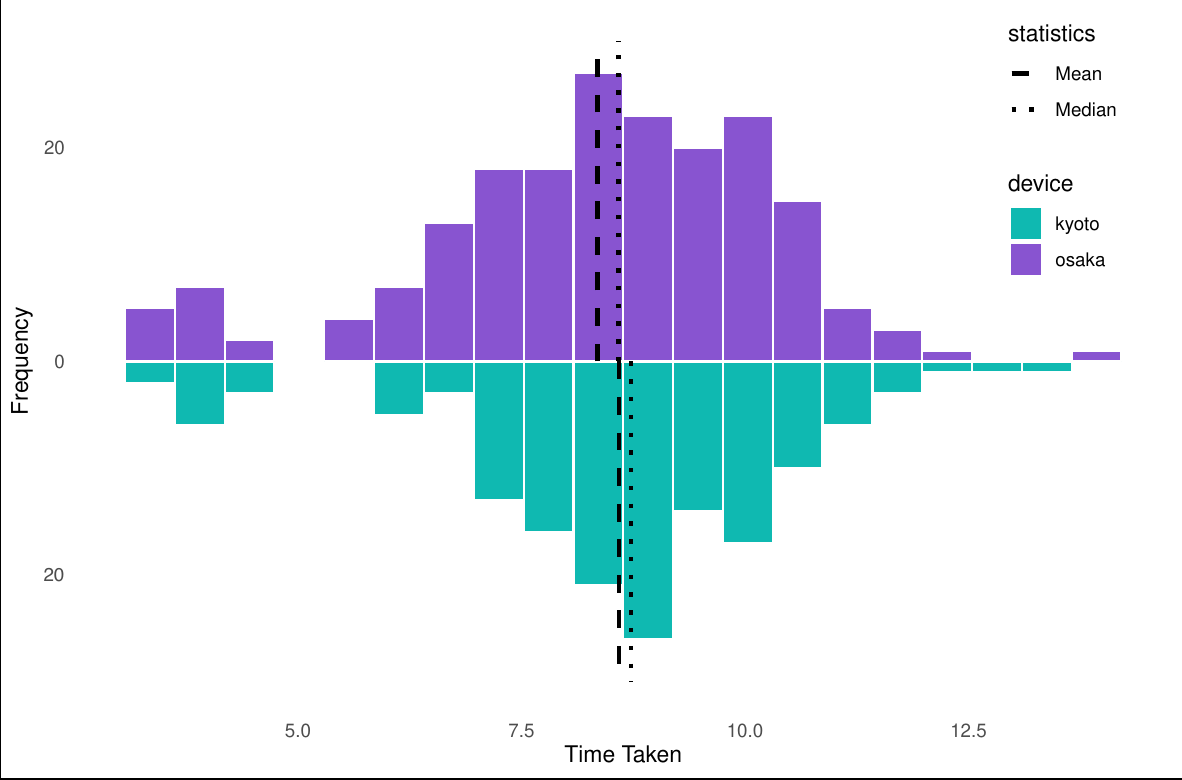}
    \caption{Histograms of execution time (in seconds) statistics for quantum circuits on different quantum computers}
    \label{fig:combined quantum computers}
\end{figure}

We follow a similar approach as in \ref{sec:simulator-time} to analyze the execution times of the circuits on real quantum computers.
Table~\ref{tab:execution time on quantum computers} and Fig.~\ref{fig:combined quantum computers} present the distributions of the execution times of the quantum circuits on the two studied quantum computers. Careful readers may notice that the average execution time on the quantum computers is longer than that on the simulators. We would like to clarify that, in Figs.~\ref{fig:combined simulators} and ~\ref{fig:combined quantum computers}, for both simulators and real devices, "execution time" denotes the backend-side system (quantum) time reported by IBM. It measures the time when the QPU is dedicated to a job and does not include queuing or host-side overhead\footnote{\url{https://quantum.cloud.ibm.com/docs/guides/max-execution-time}}. On real devices this duration includes hardware and control latencies, in particular qubit measurement and the per-shot repetition delay, which accumulate on the device side\footnote{\url{https://quantum.cloud.ibm.com/docs/guides/repetition-rate-execution}; \url{https://qiskit-community.github.io/qiskit-experiments/stable/0.6/manuals/measurement/restless_measurements.html}}. Simulators execute the same circuits in software and do not incur these hardware latencies, so in our workload their measured durations can be shorter.

\textbf{The execution times of quantum circuits on quantum computers varies significantly across different circuits.}
As shown in Table~\ref{tab:execution time on quantum computers} and Fig.~\ref{fig:combined quantum computers}, for the 192 quantum circuits executed at \texttt{ibm\_osaka} and the 148 at \texttt{ibm\_kyoto}, both the maximum and minimum execution times are approximately 13 seconds and 3 seconds, respectively, with both average and median times around 8 or 9 seconds. This indicates that the actual execution times of different quantum circuits on quantum computers can vary significantly. Such variations are challenging to estimate accurately with current methods, specifically the IBM Quantum Platform. Accurately predicting the execution time of quantum circuits on quantum computers is crucial, underlining the necessity and innovation of the predictive model proposed in this paper.

\begin{table}[h]
\centering
\caption{Statistical information on quantum circuit execution times (in seconds) on quantum computers}
\label{tab:execution time on quantum computers}
\begin{tabular}{@{}lcc@{}}
\toprule
\textbf{Statistic} & \textbf{\texttt{ibm\_osaka}} & \textbf{\texttt{ibm\_kyoto}} \\ \midrule
Count & 192 & 148 \\
Mean & 8.35 & 8.59 \\
Standard Deviation & 1.93 & 1.88 \\
Minimum & 3.11 & 3.25 \\
25\% & 7.23 & 7.70 \\
50\% & 8.58 & 8.72 \\
75\% & 9.72 & 9.83 \\
Maximum & 13.70 & 13.38 \\ \bottomrule
\end{tabular}
\end{table}

\textbf{The execution times of the same circuits on different quantum computers are similar and highly correlated.}
Due to the varying number of quantum circuits executed on different quantum computers (\texttt{ibm\_osaka} hosted 192 while \texttt{ibm\_kyoto} had 148), we initially filtered the quantum circuits to analyze only those that were executed on both \texttt{ibm\_osaka} and \texttt{ibm\_kyoto}. In fact, all the quantum circuits executed on \texttt{ibm\_kyoto} were also executed on \texttt{ibm\_osaka}, thus providing us with execution times data for a total of 148 quantum circuits on both computers for our analysis.

Similarly, we employ the Shapiro-Wilk test to assess whether the execution times of quantum circuits on \texttt{ibm\_osaka} and \texttt{ibm\_kyoto} adhere to a normal distribution.
The outcomes of the Shapiro-Wilk Test are detailed in the first two rows of Table~\ref{tab:statistical test results on different quantum computers}, where the $p$-values are significantly less than 0.05, indicating that the execution times of quantum circuits on both \texttt{ibm\_osaka} and \texttt{ibm\_kyoto} do not conform to a normal distribution.

Considering the non-normal distribution of the execution times on the two quantum computers, similar to comparing the execution times of quantum circuits on simulators, we apply the Wilcoxon Signed-Rank Test to determine if there is a significant difference in the distribution of execution times across the two quantum computers.
As illustrated in the third row of Table~\ref{tab:statistical test results on different quantum computers}, the $p$-value is 0.710, which exceeds 0.05. Thus, we accept the null hypothesis, indicating no statistically significant difference between the distributions. This suggests that the execution times for the same quantum circuits on \texttt{ibm\_osaka} and \texttt{ibm\_kyoto} are similar.

Lastly, the results of the Spearman's Rank Correlation Coefficient measurement, presented in the fourth row of Table~\ref{tab:statistical test results on different quantum computers}, show a $p$-value significantly below 0.05 and a $\rho$ value of 0.880. Thus we establish that there is a strong positive correlation in the execution times of quantum circuits on these two quantum computers. Generally, this indicates that for the same quantum circuit, if the execution time is long on one quantum computer, it tends to be long on the other, and vice versa, supporting the conclusion that the execution times on \texttt{ibm\_osaka} and \texttt{ibm\_kyoto} do not significantly differ.

\begin{table}[h]
\centering
\caption{Statistical test results for comparing the circuits' execution times on different quantum computers}
\label{tab:statistical test results on different quantum computers}
\begin{tabular}{l|c}
\hline
\textbf{Test} & \textbf{Results} \\ \hline
Shapiro-Wilk Test on \texttt{ibm\_osaka} & $p$-value $= 1.014 \times 10^{-5}$ \\
Shapiro-Wilk Test on \texttt{ibm\_kyoto} & $p$-value $= 5.176 \times 10^{-5}$ \\
Wilcoxon Signed-Rank Test & $p$-value $= 0.710$ \\
Spearman's Rank Correlation Coefficient & $p$-value $< 2.2 \times 10^{-16}$; $\rho = 0.880$ \\
\hline
\end{tabular}
\end{table}

Similarly, we examine how simple, target-independent features relate to execution time on quantum computers. On \texttt{ibm\_osaka} and \texttt{ibm\_kyoto}, Table~\ref{tab:corr-qc} shows that \emph{depth} is weak both monotonically and linearly: \texttt{ibm\_osaka} ($\rho=0.246$, $r=0.220$) and \texttt{ibm\_kyoto} ($\rho=0.221$, $r=0.231$). \emph{Num\_qubits} shows moderate association in both senses: \texttt{ibm\_osaka} ($\rho=0.658$, $r=0.571$) and
\texttt{ibm\_kyoto} ($\rho=0.556$, $r=0.471$). \emph{Two-qubit gate count} has weak-to-moderate monotonic and weak linear association with time: \texttt{ibm\_osaka} ($\rho=0.355$, $r=0.291$) and \texttt{ibm\_kyoto} ($\rho=0.275$, $r=0.212$). \emph{Num\_gates} shows weak to modest monotonic and linear associations: \texttt{ibm\_osaka} ($\rho=0.201$, $r=0.242$) and \texttt{ibm\_kyoto} ($\rho=0.309$, $r=0.369$). Taken together, the weak linear relations of individual features indicate that single features are insufficient as predictors, so we use learned models that combine features; in addition, the differences between backends suggest that backend dependent features and models are needed.
\begin{table}[h]
\centering
\caption{Monotonic vs.\ linear associations with execution time on quantum computers.}
\label{tab:corr-qc}
\begin{tabular}{lrrrr}
\toprule
& \multicolumn{2}{c}{Spearman $\rho$} & \multicolumn{2}{c}{Pearson $r$} \\
\cmidrule(lr){2-3}\cmidrule(lr){4-5}
Feature & \textbf{\texttt{ibm\_osaka}} & \textbf{\texttt{ibm\_kyoto}} & \textbf{\texttt{ibm\_osaka}} & \textbf{\texttt{ibm\_kyoto}} \\
\midrule
Depth           & 0.246 & 0.221 & 0.220 & 0.231 \\
Num\_qubits           & 0.658 & 0.556 & 0.571 & 0.471 \\
Two-qubit gate count  & 0.355 & 0.275 & 0.291 & 0.212 \\
Num\_gates           & 0.201 & 0.309 & 0.242 & 0.369                \\
\bottomrule
\end{tabular}
\end{table}

\subsubsection{Evaluation of the accuracy of execution times estimated by the IBM Quantum Platform}

Given that quantum computing resources are limited and expensive, it underscores the necessity of accurately predicting the execution times of quantum circuits on quantum computers. The IBM Quantum Platform provides an estimation of the quantum execution time when a quantum circuit is submitted for execution. It is important to note that the execution‑time estimates provided by the IBM Quantum Platform represent worst‑case values for the circuit execution itself and do not include any queueing delays. We first observe whether the estimtation provided by the IBM Quantum Platform meets our needs. When we submit a quantum circuit to a quantum computer provided by the IBM Quantum Platform for execution, we can obtain the predicted execution time of the quantum circuit by sending a GET request\footnote{https://docs.quantum.ibm.com/api/runtime/tags/jobs}. In this way, we recorded the estimated execution times of 192 quantum circuits executed at \texttt{ibm\_osaka} and 148 at \texttt{ibm\_kyoto}, provided by the IBM Quantum Platform. 
We then compare the estimated execution times with the actually measured execution time using the metrics described in \ref{sec:eval-metrics}. The results are presented in Table~\ref{tab:Evaluation of IBM Quantum Platform Prediction Results}.

\begin{table}[h!]
  \centering
  \caption{Evaluation of IBM Quantum Platform's execution time predictions}
  \begin{tabular}{ccc}
    \hline
    \textbf{MSE} & \textbf{R-squared} & \textbf{NMSE} \\
    \hline
    39981.699 & -10978.628 & 10979.628 \\
    \hline
  \end{tabular}
  \label{tab:Evaluation of IBM Quantum Platform Prediction Results}
\end{table}

\textbf{The estimation accuracy of the IBM Quantum Platform for execution times is notably inadequate.}
By observing the MSE, R-squared, and NMSE, as shown in Table~\ref{tab:Evaluation of IBM Quantum Platform Prediction Results}, we realize that the IBM Quantum Platform's predictive performance for execution times is very poor, since it reports a worst case circuit execution time rather than an expected value. For example, the $R^2$ value is negative, indicating that the prediction results are less accurate than simply predicting the average execution time. This necessitates the proposal of a new approach to better estimate the execution times of quantum circuits on quantum computers.

\begin{tcolorbox}[title=Highlights of RQ1]
On both simulators and real quantum computers, execution times vary substantially across circuits. Simple, target-independent features have nontrivial but limited associations with execution time, often monotonic rather than linear. These observations underscore the need for learned models that combine multiple features, while the backend-to-backend differences in the associations motivate the consideration of backend-dependent features and models. Accordingly, RQ2 and RQ3 aim to develop accurate, backend-aware, model-based predictions.
\end{tcolorbox}

\subsection{RQ2: How well can we estimate the execution times of quantum circuits on simulators, and what are the important predictors?} \label{sec:RQ2}
Based on the approach described in \ref{sec:model}, we build and evaluate our model for predicting the execution times of the quantum circuits on simulators. Same as in RQ1, we consider all the 1,510 selected circuits and their execution times on the two studied simulators: \texttt{FakeWashington} and \texttt{FakeSherbrooke}.

\subsubsection{The overall performance of the model in predicting the execution times of quantum circuits on simulators} 
Based on the evaluation methods and metrics described in Section~\ref{sec:evaluatino}, we evaluate the performance of our model in predicting the execution times of quantum circuits on simulators using a test set comprising 10\% of the entire dataset. The test set, with 302 data instances, is relatively large to provide a robustness evaluation of the model performance. Table~\ref{tab:prediction results on simulators} shows the evaluation results.

\begin{table}[ht]
\centering
\caption{Evaluation of our model's prediction results on simulators}
\begin{tabular}{cccc}
\toprule
\multicolumn{2}{c}{\textbf{Test Data Points:} 302} & \multicolumn{2}{c}{\textbf{Simulators:} \texttt{FakeWashington}, \texttt{FakeSherbrooke}} \\
\midrule
\textbf{Model} & \textbf{MSE} & \textbf{R-squared} & \textbf{NMSE} \\
\midrule
Full Features & \textbf{0.081} & \textbf{0.943} & \textbf{0.057} \\
Graph Features Only & 0.103 & 0.928 & 0.072 \\
Global Features Only & 0.610 & 0.573 & 0.427 \\
\texttt{FakeWashington} Only & 0.090 & 0.912 & 0.088 \\
\texttt{FakeSherbrooke} Only & 0.335 & 0.896 & 0.104 \\
\bottomrule
\end{tabular}
\label{tab:prediction results on simulators}
\end{table}

In Table~\ref{tab:prediction results on simulators}, the \textbf{Full Features} row shows the results when all the features (both global features and graph features) are used to train the model.
The \textbf{Graph Features Only} and \textbf{Global Features Only} rows indicate the results of the model assessment when we only use the graph features and the global features, respectively, for training the model.
The last two rows in the table show the results when we trained and evaluated the models using only the data corresponding to the simulators \texttt{FakeWashington} and \texttt{FakeSherbrooke}, respectively, using all the features. 

\textbf{Our model predicts the execution times of quantum circuits on simulators with high accuracy.}
As shown in Table~\ref{tab:prediction results on simulators}, our model achieves an $R^2$ score of 0.943 when considering all the features and combining the data for the two simulators. 
As expected, there is a strong correlation among MSE, $R^2$, and NMSE: smaller values of MSE and NMSE correspond to a larger $R^2$ score. 
We note significant deterioration in model performance when we only use the global features, indicating the critical role of the graph features. Our finding reveals that \textbf{the structure of a quantum circuit (e.g., how the qubits and gates are connected) plays a more important role than the global features (e.g., the number of qubits) in predicting the circuit's execution time on simulators}. It could be partially because the structure information (i.e., the graph features) already implicitly captures the global information: for example, the number of qubits can be inferred from the circuit structure itself. Nevertheless, combining both the graph features and the global feature still performs better than using the graph graph feature alone.
We also find that the model does not perform as well on individual simulator datasets (FakeWashington and FakeSherbrooke) compared to the full dataset, with $R^2$ values of 0.912 and 0.896, respectively. This indicates that the relationships between the execution times and the features are consistent between the two simulators; thus, a larger training dataset can lead to better performance.

Below, we look into the importance of the detailed features.

\subsubsection{Assessment of the importance of different components of the graph features}\label{sec:sim-graph-features}

To assess the importance of different components of the graph features, we retained all the global features but separately removed components of the graph features: Node Type, Qubit Index, T1, T2, and Node Index, retraining and evaluating the model after each removal. The outcomes from the model evaluations using MSE, R-squared, and NMSE are correspondingly labeled as \textbf{Without Node Type Feature}, \textbf{Without Qubit Index Feature}, \textbf{Without T1, T2 Feature}, and \textbf{Without Node Index Feature}, respectivelly, as shown in Table~\ref{tab:Evaluation of Prediction Results on Simulators with Missing Components of the Graph Features}.

\begin{table}[ht]
\centering
\caption{Evaluation of prediction results on simulators with missing components of the graph features}
\label{tab:Evaluation of Prediction Results on Simulators with Missing Components of the Graph Features}
\begin{tabular}{@{}cccc@{}}
\toprule
\multicolumn{2}{c}{\textbf{Test Data Points:} 302} & \multicolumn{2}{c}{\textbf{Simulators:} \texttt{FakeWashington}, \texttt{FakeSherbrooke}} \\ \midrule
\textbf{Model} & \textbf{MSE} & \textbf{R-squared} & \textbf{NMSE} \\ \midrule
Without Node Type Feature  & 0.053  & 0.961  & 0.039  \\
Without Qubit Index Feature & 0.175  & 0.878  & 0.122  \\
Without T1, T2 Feature    & 0.460  & 0.678  & 0.322  \\
Without Node Index Feature & 0.097  & 0.932  & 0.068  \\ \bottomrule
\end{tabular}
\end{table}

\textbf{The T1, T2 features contribute most significantly to the accuracy of the model, whereas the Node Type feature has a negative impact on the model's performance.}
We observe a marked decline in model performance when the Qubit Index feature, the T1, T2 features, or the Node Index feature are missing, as evidenced by higher MSE and NMSE, and lower $R^2$ value. This underscores the critical role these features play in ensuring the model's accuracy.
Particularly, the performance notably deteriorates when the model lacks the T1, T2 features. The corresponding values for MSE, $R^2$ value, and NMSE are 0.460, 0.678, and 0.322, respectively. Given that T1 measures how long a qubit stays in its excited state before returning to its ground state due to energy loss, and T2 measures the duration a qubit maintains its quantum state before environmental noise causes it to lose coherence, a longer T1 or T2 time may require a simulator to take longer time to simulator the circuit behavior. This elucidates why the omission of the T1 and T2 features markedly degrades model performance.

An anomaly was observed \textbf{when the Node Type feature was omitted: the $R^2$ increases to 0.961, higher than that of the full-feature model (with a $R^2$ of 0.943)}, and the MSE and NMSE values decrease accordingly, indicating improved model performance. This suggests a negative impact of the Node Type feature on the model. This finding indicates that the type of a quantum gate does not provide more information than the T1 and T2 information; rather, adding the 46-dimension one-hot encoding for the node type significantly increases the parameters to be trained and thus decreases the model performance.

The substantial impact of the T1 and T2 features on execution time prediction can be theoretically explained by the underlying simulation mechanism in Qiskit. When using a simulator such as \texttt{FakeWashington} or \texttt{FakeSherbrooke}, Qiskit automatically constructs a noise model based on the backend's calibration snapshot\footnote{https://qiskit.github.io/qiskit-aer/tutorials}. This snapshot includes qubit-level properties such as T1 and T2, gate durations, and error rates. During simulation, T1 and T2 are used to model decoherence processes (e.g., amplitude and phase damping), which are inserted between operations or during idle time. Longer coherence times thus require the simulator to maintain precise density matrix evolution for a longer period, increasing the overall runtime. This explains the marked performance degradation observed when T1 and T2 features are excluded from the model.

\subsubsection{Assessment of the importance of different components of the global features} \label{sec:sim-global-features}

As discussed in Section~\ref{sec:model-construction}, the global features are significantly smaller in dimension compared to the graph features and each component of the global features can be subdivided into specific features (such as circuit depth, certain quantum gates, etc.), removing a single feature from the global features has a minimal impact on the model and makes the entire training and validation process overly cumbersome. 
Therefore, we utilize SHAP  (SHapley Additive exPlanations) values to assess the importance of each individual feature within the global features.

\begin{figure} [h]
    \centering
    \includegraphics[width=1\linewidth]{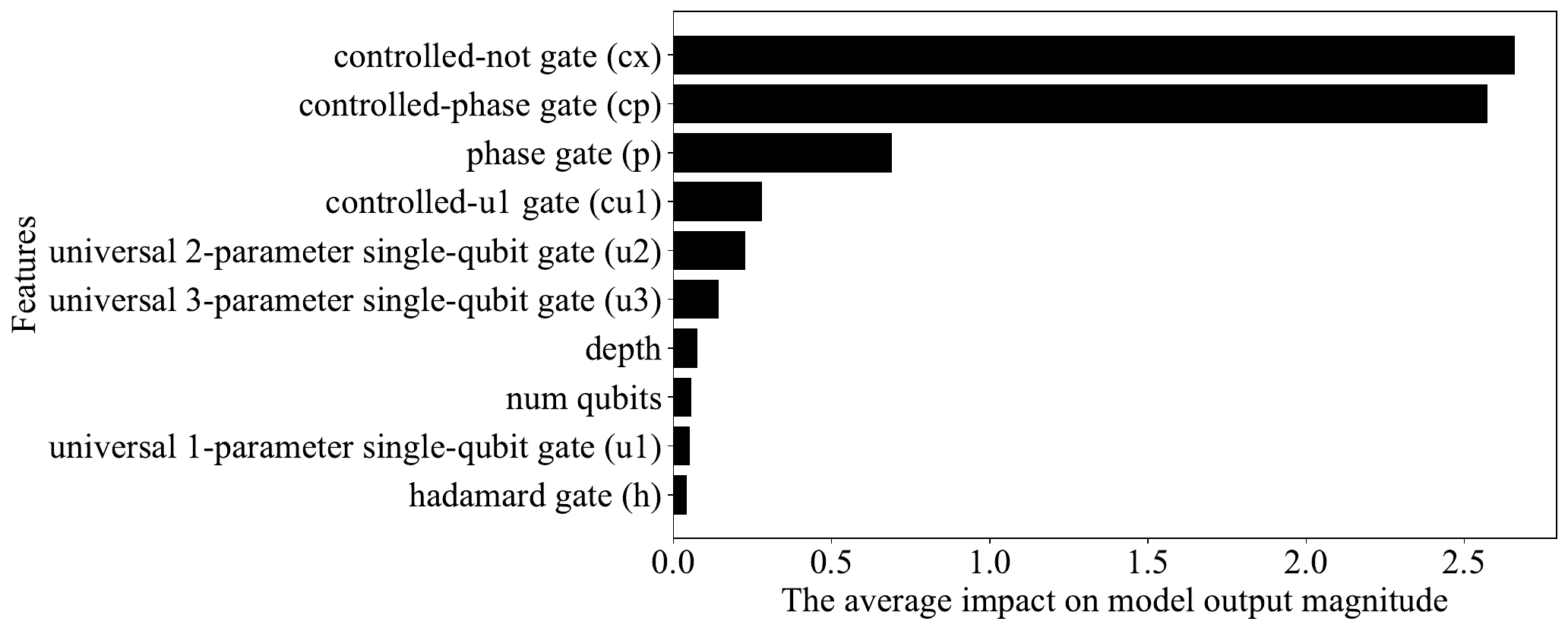}
    \caption{The mean of the absolute values of the SHAP values for different features of the global features on simulators}
    \label{fig:shap_value_simulator}
\end{figure}

SHAP values quantify the contribution of each feature to a prediction based on Shapley values from game theory, ensuring a fair distribution of feature importance. The core idea of SHAP values is to measure the impact of a feature on the model's prediction by considering the difference in the prediction when the feature is present versus when it is absent. By calculating the contribution of each feature value and averaging these contributions across all possible combinations of features, SHAP values provide a fair measure of each feature's importance~\cite{lundberg2017unified}.
A positive SHAP value indicates that the feature increases the predicted value, whereas a negative SHAP value suggests that the feature decreases the predicted value. The larger the absolute value of a SHAP value, the greater its impact on the model's predicted value, and correspondingly, the higher the contribution of the corresponding feature to the model. Therefore, on the test set, for each feature in the global features, we calculate the average of the absolute values of the SHAP values predicted for each data point. These averages are then sorted from largest to smallest, with larger values indicating more important features. The results are visualized in Fig.~\ref{fig:shap_value_simulator}.

\textbf{The number of gates, particularly two-qubit gates such as the Controlled-Not (CX) and Controlled-Phase (CP) gates, plays the most important roles among the global features in predicting the execution times of quantum circuits.}
As shown in Fig.~\ref{fig:shap_value_simulator}, due to the large number of features in the global features (41), we only display the top 10 most important features. From Fig.~\ref{fig:shap_value_simulator}, it can be observed that, apart from the depth and number of qubits features, the other important features are various quantum gates. The two most significant gates are the CX and CP gates, both of which are two-qubit gates. This observation aligns with physical and compilation characteristics of quantum circuits. These gates are very common in quantum circuits, and it can be inferred that their prevalence also reflects the complexity of a quantum circuit, thereby playing a significant role in predicting the execution time of the circuit. %In particular, two-qubit gates are known to have longer gate time (the time taken for a gate operation) than one-qubit gates~\cite{tomesh2022supermarq}.
Two-qubit gates are significantly more expensive to simulate than single-qubit gates because they can create entanglement between qubits, thereby increasing the complexity of the quantum state the simulator must track.
Moreover, these two-qubit gates often undergo nontrivial decomposition during compilation. For example, a CP gate is typically decomposed into a combination of basis gates including multiple CX gates and single-qubit rotations. On \texttt{FakeWashington}, the CX gate is the native two-qubit gate; however, on \texttt{FakeSherbrooke}, the native entangling gate is ECR gate, meaning every CX gate in the circuit is further decomposed into multiple ECR gates and surrounding rotations. As a result, circuits containing high-level two-qubit gates—especially on non-CX-native devices—undergo deeper compilation and produce a larger number of physical-level operations for simulation.
This explanation is quantitatively supported by Table~\ref{tab:avg_gate_times}, which lists average gate durations on both simulators. The native two-qubit gates, CX (550.41~ns) and ECR (533.83~ns), are nearly an order of magnitude slower than the single-qubit gates SX and X (35.56–56.89~ns). RZ gates incur almost no time cost. This stark difference explains why two-qubit gates dominate simulation time prediction: not only are they more expensive to simulate per gate, but they also propagate noise, entanglement, and depth, further burdening the simulator—especially when their number is inflated due to compilation and qubit topology constraints.

\begin{tcolorbox}[title=Highlights of RQ2]
Our graph transformer-based model, based on the structure (graph) and global features of quantum circuits, can accurately predict the execution times of quantum circuits on simulators, achieving an $R^2$ score of 0.943 (the $R^2$ further increases to 0.961 when the \textit{Node Type} feature is excluded). Overall, the graph features play a more important role than the global features; among the individual features, the \textit{T1, T2} features within the graph features, and the number of different types of two-qubit gates (e.g., CX gates) within the global features, are among the most important features.
\end{tcolorbox}

\subsection{RQ3: How well can we estimate the execution times of quantum circuits on real quantum computers, and what are the important predictors?} \label{sec:RQ3}

Similar to RQ2 (\ref{sec:RQ2}), we build and evaluate our model for predicting the execution times of the quantum circuits on real quantum computers, based on the approach described in \ref{sec:model}. Same as in RQ1, we consider the 340 selected circuits that are selected to execute on real computers: 192 circuits on \texttt{ibm\_osaka} and 148 on \texttt{ibm\_kyoto}. The details for the circuit selection through active learning are described in \ref{sec:active-learning}.

\subsubsection{The overall performance of the model in predicting the execution times of quantum circuits on quantum computers}

Similar to RQ2, in a 10-fold cross-validation, we evaluate the performance of our model for predicting the execution times of the quantum circuits on real quantum computers. 
Table~\ref{tab:prediction results on quantum computers} shows the evaluation results.
Similar to RQ2, we first evaluated the original model using all the features and data for both computers. We then retrained and assessed the models' performance using only the graph features and global features individually. Finally, we utilized all the features, but specifically trained and evaluated the model with data exclusively from the quantum computers \texttt{ibm\_osaka} and \texttt{ibm\_kyoto}, employing 10-fold cross-validation for each.
\textbf{With a relatively small dataset, our approach can accurately predict the execution times of quantum circuits on real quantum computers, achieving an $R^2$ score of 0.905.}
As shown in Table~\ref{tab:prediction results on quantum computers}, the models' performance, measured in terms of Average MSE, Average NMSE, and Average $R^2$ values, are not as good as that of the simulators (RQ2).
However, a $R^2$ larger than 0.9 still indicates a high precision of the prediction results. 
Nevertheless, our results indicate that execution times of quantum circuits on quantum computers are more challenging to predict than on simulators, potentially due to 1) the greater variability and more unpredictable disturbances in quantum computers and 2) the challenge of collecting a large amount of data.  
Additionally, we find that the MSE for predicting the execution times of quantum circuits on quantum computers is significantly higher. This is because the execution times of quantum circuits on quantum computers are substantially longer than their run times on simulators, making a higher MSE reasonable and justifiable. This highlights the importance of calculating Normalized Mean Squared Error, as we see that the NMSE results are similar in Table~\ref{tab:prediction results on simulators} and Table~\ref{tab:prediction results on quantum computers}.

Interestingly, \textbf{the graph features and global features are almost equally capable of predicting the execution times}. This observation is different from that from RQ2: on simulators, graph features play a significantly more important role. This difference may be because simulators are more sensitive to the complexity of the circuit structure: a more complex structure may take more time to simulate the circuit behavior; however, a real quantum computer is natively capable of handling complex quantum circuit structures.

\begin{table}[ht]
\centering
\caption{Evaluation of prediction results on quantum computers}
\begin{tabular}{cccc}
\toprule
\multicolumn{2}{c}{\textbf{Data Points:} 340} & \multicolumn{2}{c}{\textbf{Computers:} \texttt{ibm\_osaka}, \texttt{ibm\_kyoto}} \\
\midrule
\textbf{Model} & \textbf{Average MSE} & \textbf{Average R-squared} & \textbf{Average NMSE} \\
\midrule
Full Features & 0.305 & 0.905 & 0.095 \\
Graph Features Only & 0.325 & 0.900 & 0.100 \\
Global Features Only & 0.346 & 0.894 & 0.106 \\
\texttt{ibm\_osaka} Only & \textbf{0.239} & \textbf{0.927} & \textbf{0.073} \\
\texttt{ibm\_kyoto} Only & 0.405 & 0.833 & 0.167 \\
\bottomrule
\end{tabular}
\label{tab:prediction results on quantum computers}
\end{table}

When exploring the model's predictive performance on datasets corresponding to individual quantum computers (\texttt{ibm\_osaka} or \texttt{ibm\_kyoto}), we find that the model performs better on \texttt{ibm\_osaka} compared to the overall data or the \texttt{ibm\_kyoto} data, leading to more accurate predictions. In addition, the result indicates that different quantum computers may exhibit slightly different relationships between the execution times and the features.

\subsubsection{Assessment of the importance of different components of the graph features}
\label{sec:computer-graph-features}

Similar to \ref{sec:sim-graph-features}, we iteratively removed parts of the graph features and performed 10-fold cross-validation to evaluate the model performance using the remaining features, with the results presented in Table~\ref{tab:Evaluation of Prediction Results on Quantum Computers with Missing Components of the Graph Features}.

\textbf{All features within the graph features positively impact the model's performance; however, none of the graph features are critical, as removing any one of these features does not significantly impact the model's performance. }
In particular, unlike the results of RQ2, the T1, T2 features do not play a significantly important role: in fact, the absence of the T1, T2 features results in the smallest increases in Average MSE and Average NMSE, and the least decrease in Average R-squared, which contrasts sharply with their highest importance for predicting execution times on simulators in RQ2. As explained in RQ2, a longer T1 or T2 time may require a simulator to take a longer time to simulate the circuit behavior; however, when a circuit is executed in a real quantum computer, intuitively, the execution time does not depend on the T1, T2 times: %T1, T2 times control the maximum time a quantum circuit can execute reliably, not the actual execution time. These differences can be theoretically attributed to the execution model of real quantum hardware. 
Unlike noise-aware simulators that explicitly simulate decoherence based on T1 and T2 values, real quantum computers only rely on pulse calibration and gate scheduling, where T1 and T2 represent the upper bounds on coherent operation windows, rather than determining the actual gate time. Therefore, T1 and T2 do not directly affect the execution time of a circuit, which instead depends on gate durations, pulse alignment, and scheduling constraints on hardware. This explains why removing the T1, T2 features does not significantly degrade performance in this setting.
\textbf{Our observations highlight the intrinsic differences between modeling quantum circuits' execution times on real quantum computers and simulators.}

\begin{table}[ht]
\centering
\caption{Evaluation of prediction results on quantum computers with missing components of the graph features}
\begin{tabular}{cccc}
\toprule
\multicolumn{2}{c}{\textbf{Data Points:} 340} & \multicolumn{2}{c}{\textbf{Computers:} \texttt{ibm\_osaka}, \texttt{ibm\_kyoto}} \\
\midrule
\textbf{Model} & \textbf{Average MSE} & \textbf{Average R-squared} & \textbf{Average NMSE} \\
\midrule
Without Node Type Feature & 0.322 & 0.900 & 0.100 \\
Without Qubit Index Feature & 0.335 & 0.897 & 0.103 \\
Without T1, T2 Feature & 0.315 & 0.903 & 0.097 \\
Without Node Index Feature & 0.337 & 0.897 & 0.103 \\
\bottomrule
\end{tabular}
\label{tab:Evaluation of Prediction Results on Quantum Computers with Missing Components of the Graph Features}
\end{table}

\subsubsection{Assessment of the importance of different components of the global features}

\begin{figure} [h]
    \centering
    \includegraphics[width=1\linewidth]{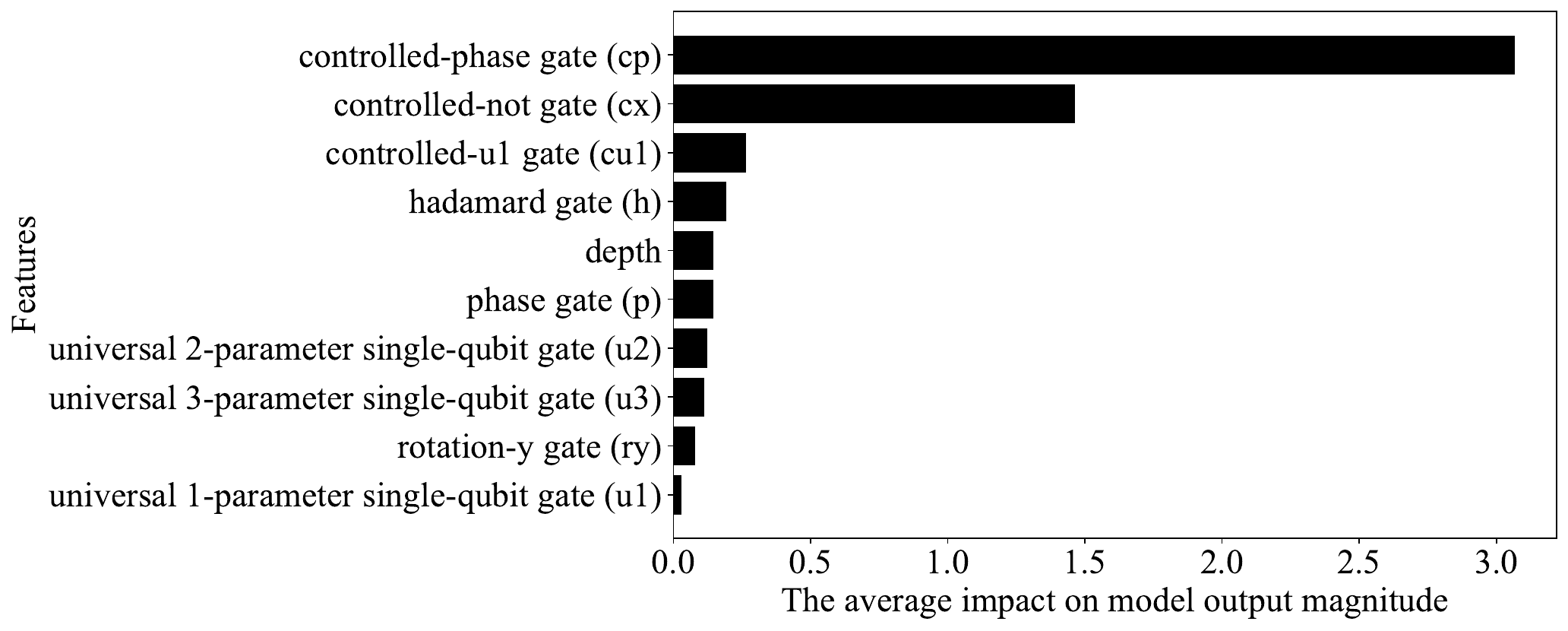}
    \caption{The mean of the absolute values of the SHAP values for different features of the global features on quantum computers}
    \label{fig:shap_value_computer}
\end{figure}

Similar to RQ2, we use SHAP values to assess the importance of different features in the global features set for predicting execution times on quantum computers. We calculate the average of the absolute values of the SHAP values obtained from 10-fold cross-validation, and then average these values to obtain the final SHAP value, as shown in Fig.~\ref{fig:shap_value_computer}. 

\textbf{The numbers of different types of gates, particularly two-qubit gates such as CP and CX gates, remain the most crucial features among the global features in predicting the execution times of quantum circuits on quantum computers.} 
We find that, compared to those identified in the model on the simulators, the top 10 most important features among the global features on quantum computers lack the number of qubits feature, but include the rotation-y gate feature. The remaining features remain the same. This indicates that, whether on simulators or on quantum computers, the quantity of common gates within quantum circuits plays a crucial role in accurately predicting the execution times of these circuits. In particular, despite the swapping of their rankings, the numbers of two-qubit gates (e.g., CP gate, CX gate) still play the most important roles.%, due to the fact that two-qubit gates have significantly longer gate times than single-qubit gates~\cite{tomesh2022supermarq}. 
This is consistent with the compilation behavior on real quanutm devices. For example, both \texttt{ibm\_osaka} and \texttt{ibm\_kyoto} support only a limited native gate set, specifically \texttt{[ecr, id, rz, sx, x]}, and lack native CX or CP support. As a result, gates like CX and CP must be decomposed into one or more ECR gates combined with single-qubit rotations. This decomposition increases the number of scheduled instructions and total circuit duration, leading to longer execution times—particularly for circuits with many high-level two-qubit gates.

Many previous studies have emphasized the importance of T gates in quantum cost analysis, such as~\cite{rahman2022grover}. However, it is noteworthy that in this paper, the importance of the T gate is relatively low in both execution time estimation on simulators and on real quantum computers. Specifically, the T gate does not appear among the top-10 most important global features as determined by our feature importance analysis.
Upon further inspection of the dataset used in this work, which consists of 1,510 quantum circuits, we find that only 104 out of 1,510 circuits contain any T gates, and the number of T gates in those circuits is generally small. This limited presence likely explains why the T gate does not have a significant impact on execution time prediction in our model. As a result, although the T gate may play a crucial role in other contexts, it does not contribute substantially to performance estimation in our setting.

\begin{tcolorbox}[title=Highlights of RQ3]
With a relatively small dataset, our approach can also accurately predict the execution times of quantum circuits on real quantum computers, achieving a $R^2$ score of 0.905 using the combined data from different quantum computers and a $R^2$ score of up to 0.927 when using the data from a single computer. Different from that for the simulators, the graph features and global features are almost equally capable of predicting the execution times on real computers. Our results highlight the intrinsic differences between modeling the execution times of quantum circuits executed on simulators and real quantum computers.
\end{tcolorbox}
\section{Discussion}
\label{sec:discussion}
In RQ3 (\ref{sec:RQ3}), we build and evaluate our model for predicting the execution times of the quantum circuits on real quantum computers. Although the selected dataset contains quantum circuits with qubit counts ranging from 2 to 127, we note that circuits generated from the same algorithm family can still exhibit similar structural patterns, even if their qubit counts differ. Therefore, using a standard cross-fold validation on the dataset where many circuits originate from the same algorithm family may result in significant overlap between the training and test sets, potentially hindering the evaluation of the model's generalization capability.

To address this, we propose a new cross-validation strategy that partitions the dataset by algorithm families. Specifically, in each fold, we select one algorithm family as the test set and use circuits from the remaining algorithm families as the training set. We observe that the selected 340 circuits span 10 distinct algorithm families; however, one algorithm family contains only two circuits. To ensure sufficient representation, we merge this algorithm family with the algorithm family containing the second fewest (26) circuits, resulting in a total of 9 folds for cross-validation.

Consequently, the number of circuits in each fold is \mbox{[40, 46, 30, 35, 38, 28, 47, 40, 36]}.  
The prediction performance obtained with this algorithm-family-based cross-validation scheme is summarised in Table~\ref{tab:prediction-results-new-cv}.

\begin{table}[ht]
\centering
\caption{Prediction performance under the proposed algorithm-family-based cross-validation (9 folds)}
\begin{tabular}{cccc}
\toprule
\multicolumn{2}{c}{\textbf{Data Points:} 340} & \multicolumn{2}{c}{\textbf{Computers:} \texttt{ibm\_osaka}, \texttt{ibm\_kyoto}} \\
\midrule
\textbf{Model / Setting} & \textbf{Average MSE} & \textbf{Average R-squared} & \textbf{Average NMSE} \\
\midrule
Full Features & 0.485 & 0.878 & 0.122 \\
\bottomrule
\end{tabular}
\label{tab:prediction-results-new-cv}
\end{table}

Compared with the results in Section~\ref{sec:computer-graph-features}, the algorithm-family-partitioned cross-validation reveals notable differences: the average $R^{2}$ score falls from 0.905 to 0.878, the average MSE increases from 0.305 to 0.485, while the average NMSE rises from 0.095 to 0.122.

This indicates that when the training and test sets contain circuits originating from the same algorithm family, those circuits, despite large variations in qubit count, still share strong structural similarities that
artificially improve prediction accuracy.
Hence a random split yields an optimistic assessment of performance. After restricting each algorithm family to a single fold, the model faces a more diverse structural distribution, the prediction task becomes harder, and the metrics deteriorate. This comparison confirms that an algorithm-family-based split is essential for evaluating the true generalisation capability of the model.

Future work could explore representation-learning techniques that encourage structure-invariant features, for example by adding contrastive losses that penalise predictions driven by superficial circuit patterns.

Another aspect worth discussing is the use of simulator-derived execution times as additional features for predicting real-device performance. For the circuits in our dataset with measured execution times on both the simulators and real quantum computers, we measure the Spearman's rank correlation between each pair of simulators and real quantum computers. Our correlation study (Table~\ref{tab:correlation_results}) shows that, although simulator and hardware run times differ in absolute terms, their \emph{relative} ordering is highly consistent: Spearman’s rank coefficients range from~0.788 to~0.858 with \(p<2.2\times10^{-16}\).
This strong monotonic relationship indicates that simulator timings already provide reliable ranking signals for workload selection and scheduling.
Leveraging these signals in hybrid predictive models, perhaps together with simulator back-ends calibrated to capture hardware-specific properties, is therefore a promising direction for future work.

\begin{table}[h]
\centering
\caption{Spearman's rank correlation between the circuits' execution times on simulators and real quantum computers}
\label{tab:correlation_results}
\begin{tabular}{l|c}
\hline
\textbf{Spearman's Rank Correlation Coefficient} & \textbf{Results} \\ \hline
\texttt{FakeWashington} vs \texttt{ibm\_osaka} & $p$-value $< 2.2 \times 10^{-16}$; $\rho = 0.825$ \\
\texttt{FakeWashington} vs \texttt{ibm\_kyoto} & $p$-value $< 2.2 \times 10^{-16}$; $\rho = 0.788$ \\
\texttt{FakeSherbrooke} vs \texttt{ibm\_osaka} & $p$-value $< 2.2 \times 10^{-16}$; $\rho = 0.858$ \\
\texttt{FakeSherbrooke} vs \texttt{ibm\_kyoto} & $p$-value $< 2.2 \times 10^{-16}$; $\rho = 0.809$ \\
\hline
\end{tabular}
\end{table}
\section{Threats to Validity}
\label{sec:threats-to-validity}
\subsection{Internal Validity}
When executing quantum circuits on quantum computers, the scarcity of quantum computer resources and the lengthy total time required for executing each circuit (including queuing time which usually takes hours for a circuit), prevent us from executing all the quantum circuits on quantum computers. Instead, we selected a subset of the quantum circuits with 95\% confidence for execution and data collection. We used active learning to maximize the diversity of the sampled circuits. However, this may result in a deviation between the observed results and the actual results: for example, the performance of our model may increase if more quantum circuits are considered.

\subsection{External Validity}
This study is limited to predicting the execution times of quantum circuits on simulators and quantum computers provided by the IBM Quantum Platform. It does not extend to quantum computers provided by other platforms such as Rigetti, D-Wave, or Google Quantum AI. Quantum computers developed by different platforms may require adjustments to our model when predicting the execution time of quantum circuits on the corresponding quantum computers, potentially leading to different results.

\subsection{Construct validity}
The measurement of quantum circuits' execution time may be subject to noise caused by the instability of the computing environment. For example, the purity of the states of the qubits and the entanglement with the noisy environments may influence the execution times of quantum circuits on the quantum computers. To address the measurement noise issue, we executed each circuit three times to ensure a high confidence level of the measurement. However, some circuits (fewer than 10\%) may face higher instability that needs more measurements. We chose to measure the execution time of each circuit three times to balance the measurement precision and the time and resource costs.
\section{Conclusion}
\label{sec:conclusion}
Quantum computing resources are rather limited, thus making judicious planning and utilization of quantum computers is important for both quantum computing service providers and users. 
In this paper, we first study the characteristics of quantum circuits' runtime on simulators and real quantum computers. Then, we introduce an innovative graph transformer-based model, utilizing the graph information and global information of quantum circuits to estimate their execution times. 
Our experiments indicate that the execution times of quantum circuits vary significantly across different circuits, and that our approach can accurately predict their execution times on both simulators and quantum computers. We observe that some characteristics of the quantum circuits, such as the numbers of different types of quantum gates, especially two-qubit gates, play important roles in predicting their execution times, yet we also highlight the intrinsic differences in modeling the execution times of quantum circuits executed on simulators and real computers.

This paper marks a significant advance in quantum computing resource management: accurate estimation of quantum computing resource requirements such as execution times is the key for optimizing the allocation and utilization of these scarce, costly resources. Our approach can be integrated into quantum computing platforms to provide a more accurate estimation of quantum execution time and be used as a reference for prioritizing quantum execution jobs. 
In addition, our findings provide insights for quantum program developers to optimize their circuits in terms of execution time consumption, for example, by prioritizing one-qubit gates over two-qubit gates.

Our study also sheds light on further explorations. 
First, future research may leverage our approach to automatically optimize quantum circuits for reduced execution time.
In addtion, future research may extend our study beyond the IBM Quantum Platform to include other emerging quantum computing technologies, such as Google's Sycamore or Rigetti, which could provide broader insights for understanding quantum execution times and our model's adaptability and efficiency across different quantum systems.

\section*{Acknowledgment}
We acknowledge the support of the Natural Sciences and Engineering Research Council of Canada (NSERC) (ALLRP 580933 - 22). We also thank IBM and the Digital Research Alliance of Canada for providing quantum computing and cluster resources for our experiments.

Cette recherche a été financée par le Conseil de recherches en sciences naturelles et en génie du Canada (CRSNG) (ALLRP 580933 - 22).

\bibliographystyle{ACM-Reference-Format}
\bibliography{quantum-exec-time}

\end{document}